\documentclass[12pt,letterpaper,a4paper]{article}
\usepackage[includeheadfoot,
marginratio={1:1,2:3},
width=500pt,
height=720pt,]{geometry}

\usepackage{amsmath}
\usepackage{amsfonts}
\usepackage{amssymb}
\usepackage{graphicx}
\usepackage{cancel}
\usepackage{empheq}
\usepackage{color}
\usepackage{hyperref}

\numberwithin{equation}{section}
\usepackage{float}
\restylefloat{table}
\restylefloat{figure}
\usepackage[utf8]{inputenc}
\usepackage{amsmath, amsthm, amssymb, amsfonts}
\usepackage[font=small, labelfont=bf]{caption}
\usepackage{amsmath, amsthm, amssymb, amsfonts}
\usepackage{multirow}
\usepackage{graphicx}
\usepackage{float}
\usepackage[numbers,sort&compress]{natbib}
\usepackage{enumerate}
\usepackage{amsmath}
\usepackage{amsfonts}
\usepackage{amssymb}
\usepackage{graphicx}
\usepackage{cancel}
\usepackage{empheq}
\usepackage{color}
\usepackage{hyperref}
\usepackage{float}
\restylefloat{table}
\restylefloat{figure}

\newcommand{\nc}{\newcommand}
\nc{\beq}{\begin{equation}}
\nc{\eeq}{\end{equation}}
\nc{\bea}{\begin{eqnarray}}
\nc{\eea}{\end{eqnarray}}
\def\IZ{\mathbb{Z}}

\def\ov{\overline}

\begin{document}
	{\hfill
		%
		arXiv:2309.08664}
	
	\vspace{1.0cm}
	\begin{center}
		{\Large
			Symplectic formulation of the type IIB scalar potential \\ with U-dual fluxes}
		\vspace{0.4cm}
	\end{center}
	
	\vspace{0.35cm}
	\begin{center}
		George K. Leontaris$^\diamond$, and Pramod Shukla$^\dagger$ \footnote{Email: leonta@uoi.gr, pshukla@jcbose.ac.in}
	\end{center}

	\vspace{0.1cm}
	\begin{center}
{$^\diamond$ Physics Department, University of Ioannina, \\
 University Campus, Ioannina 45110, Greece.\\
\vskip0.5cm 
$^\dagger$ Department of Physical Sciences, Bose Institute, Unified Academic Campus, \\ 
EN 80, Sector V, Bidhannagar, Kolkata 700091, India.}
	\end{center}
	
	\vspace{1cm}
	
\abstract{We present a symplectic formulation of the $N =1$ four-dimensional type IIB scalar potential arising from a flux superpotential which has four S-dual pairs of fluxes demanded by the U-dual completion arguments. Our symplectic formulation presents a very compact and concise way of expressing the generic scalar potential in just a few terms via using a set of symplectic identities along with the so-called ``axionic-flux" combinations. We demonstrate the utility of our symplectic master-formula by considering an underlying four-dimensional type IIB supergravity model based on a ${\mathbb T}^6/({\mathbb Z}_2 \times {\mathbb Z}_2)$ orientifold, in which the scalar potential induced by the U-dual flux superpotential results in a total of 76276 terms involving 128 flux parameters. Given that our symplectic formulation does not need the information about the metric of the internal background, it is applicable to the models beyond the toroidal compactifications such as to those which use orientifolds of the Calabi-Yau threefolds.}

\clearpage
	
\tableofcontents


\section{Introduction}
\label{sec_intro}

In the context of superstring compactification, toroidal orientifolds have been considered as a promising toolkit to facilitate some simple and explicit computations. Despite being simple, such internal backgrounds can still support a very rich structure to include fluxes of various kinds which can be subsequently used for generic phenomenological studies related to, for example, moduli stabilization and the search of physical vacua \cite{Kachru:2003aw, Balasubramanian:2005zx, Grana:2005jc, Blumenhagen:2006ci, Douglas:2006es, Denef:2005mm, Blumenhagen:2007sm}. In this regard, compactification backgrounds supporting the so-called non-geometric fluxes have emerged as interesting playgrounds for initiating some kind of alternate phenomenological model building \cite{Derendinger:2004jn,Grana:2012rr,Dibitetto:2012rk, Danielsson:2012by, Blaback:2013ht, Damian:2013dq, Damian:2013dwa, Hassler:2014mla, Ihl:2007ah, deCarlos:2009qm, Danielsson:2009ff, Blaback:2015zra, Dibitetto:2011qs, Plauschinn:2018wbo,Damian:2023ote}. In fact, the existence of (non-)geometric fluxes can be understood to emerge from the following chain of T-dualities acting on the NS-NS three-form flux ($H_{ijk}$) of the type II supergravity theories \cite{Shelton:2005cf},
\bea
\label{eq:Tdual}
& & H_{ijk} \longrightarrow \omega_{ij}{}^k  \longrightarrow Q_i{}^{jk}  \longrightarrow R^{ijk} \,,
\eea
where $\omega_{ij}{}^k$ denotes the geometric flux, while $Q_i{}^{jk}$ and $R^{ijk}$ correspond to the so-called non-geometric fluxes. Moreover, one can further generalize the underlying background via seeking more and more fluxes which could be consistently incorporated/allowed in the four-dimensional (4D) effective theory via say the holomorphic flux superpotential. For this purpose, the successive application of a series of T- and S-dualities turns out to have a crucial role in constructing a generalized holomorphic flux superpotential \cite{Shelton:2005cf, Aldazabal:2006up,Aldazabal:2008zza,Font:2008vd,Guarino:2008ik, Hull:2004in, Kumar:1996zx, Hull:2003kr,Aldazabal:2010ef, Lombardo:2016swq, Lombardo:2017yme}. This includes, for example, the so-called $P$-flux in type IIB setting which is needed to restore the underlying S-duality broken by the presence of the non-geometric $Q$-flux. Inclusion of various kinds of such fluxes (which act as some parameters in the 4D supergravity dynamics) can facilitate a very diverse set of superpotential couplings which can be useful for numerous model building purposes \cite{Derendinger:2004jn,Grana:2012rr,Dibitetto:2012rk, Danielsson:2012by, Blaback:2013ht, Damian:2013dq, Damian:2013dwa, Hassler:2014mla, Ihl:2007ah, deCarlos:2009qm, Danielsson:2009ff, Blaback:2015zra, Dibitetto:2011qs}. However, this also induces several complexities (such as huge size of scalar potential, tadpole conditions, and Bianchi identities) in the non-geometric flux compactification based models, something which have been witnessed at many occasions \cite{Aldazabal:2008zza,Font:2008vd,Guarino:2008ik,Danielsson:2012by,Damian:2013dq, Damian:2013dwa, Blumenhagen:2013hva,Villadoro:2005cu, Robbins:2007yv, Ihl:2007ah,Shukla:2016xdy,Plauschinn:2021hkp}.

In fact, model building efforts using non-geometric fluxes have been made mostly via considering the 4D effective scalar potentials arising from the K\"ahler and super-potentials \cite{Danielsson:2012by, Blaback:2013ht, Damian:2013dq, Damian:2013dwa, Blumenhagen:2013hva, Villadoro:2005cu, Robbins:2007yv, Ihl:2007ah, Gao:2015nra}, and during the initial phase of phenomenological studies one did not have a proper understanding of the higher-dimensional origin of such 4D non-geometric scalar potentials. However, these aspects have received significant amount of attention in recent years, e.g. see \cite{Andriot:2013xca, Andriot:2011uh, Blumenhagen:2015lta, Villadoro:2005cu, Blumenhagen:2013hva, Gao:2015nra, Shukla:2015rua, Shukla:2015bca}. Moreover, most of these studies have been based on toroidal orientifolds as such setups are among the simplistic ones, and interests in extending these ideas for model building beyond the toroidal case got attention with the studies initiated in \cite{Hassler:2014mla, Blumenhagen:2015qda, Blumenhagen:2015kja,Blumenhagen:2015jva, Blumenhagen:2015xpa,  Li:2015taa, Blumenhagen:2015xpa}. However, the main obstacle in understanding the higher dimensional origin of the 4D effective potentials in the beyond toroidal cases (such as those based on the Calabi Yau (CY) orientifolds) lies in the fact that the explicit form of the metric for a CY threefold is not known, something which has been very much central to the ``dimensional oxidation" proposal of \cite{Blumenhagen:2013hva, Gao:2015nra}. 

For that purpose, the existence of some close connections between the 4D effective potentials of type II supergravities and the symplectic geometries turn out to be extremely crucial \cite{Ceresole:1995ca, D'Auria:2007ay}. In fact it has been well established that using symplectic ingredients one can simply bypass the need of knowing the CY metric in writing the 4D scalar potentials via using explicit expressions for the moduli space metric equipped with some symplectic identities \cite{Taylor:1999ii}. For example, in the simple type IIB model having the so-called RR and NS-NS flux pair $(F, H)$ \cite{Gukov:1999ya,Dasgupta:1999ss}, the generic 4D scalar potential can be equivalently derived from two routes; one arising from the flux superpotential while the other one following from the dimensional reduction of the 10D kinetic pieces, by using the period matrices and without the need of knowing CY metric \cite{Taylor:1999ii, Blumenhagen:2003vr}. This strategy was subsequently adopted for a series of type IIA/IIB models with more (non-)geometric fluxes, leading to what is called as the `symplectic formulation'  of the 4D scalar potential; for example, see \cite{Shukla:2015hpa,Blumenhagen:2015lta,Shukla:2016hyy,Shukla:2019wfo,Shukla:2019dqd,Shukla:2019akv} for type IIB case, and \cite{Gao:2017gxk,Shukla:2019wfo,Marchesano:2020uqz,Marchesano:2021gyv} for type IIA and F-theory case. 

Implementing the successive chain of T- and S-dualities, leads to a U-dual completion of the flux superpotential which has been studied in \cite{Aldazabal:2006up, Aldazabal:2008zza, Aldazabal:2010ef, Lombardo:2016swq, Lombardo:2017yme}. Focusing on a toroidal type IIB $\mathbb T^6/(\mathbb Z_2 \times \mathbb Z_2)$ orientifold model, such a U-dual completed superpotential turns out to have 128 fluxes, and leads to a huge scalar potential having 76276 terms as observed in \cite{Leontaris:2023lfc}. Moreover, following the prescription of \cite{Villadoro:2005cu, Blumenhagen:2013hva}, the various pieces of this effective scalar potential has been rewritten in \cite{Leontaris:2023lfc} using the internal metric. In the current article, we aim to present a symplectic formulation of the flux superpotential with U-dual fluxes, which also applies beyond the toroidal case, while reproducing the results of \cite{Leontaris:2023lfc} as a particular case.

The article is organized as follows: we begin with recollecting the relevant pieces of information about the generalized fluxes and the subsequently induced superpotential in Section \ref{sec_setup}. Continuing with the T-dual completion of the flux superpotential, in section \ref{sec_Udual} we present a U-dual completion of the flux superpotential via taking a symplectic approach. Section \ref{sec_potential} presents a detailed taxonomy of the scalar potential leading to a compact and concise master formula, which is subsequently demonstrated to reproduce the toroidal results as a particular case. Finally we summarise the conclusions in Section \ref{sec_conclusions}, and present the detailed expressions of all the 36 types of scalar potential pieces in the appendix \ref{sec_Appendix1}.


\section{Preliminaries}
\label{sec_setup}
The F-term scalar potential governing the dynamics of the ${N}=1$ low energy effective supergravity can be computed from the K\"ahler potential and the flux induced superpotential by considering the following well known relation,
\bea
\label{eq:Vtot}
& & V=e^{K}\Big(K^{I\bar J}D_I W\, D_{\bar J} \ov W-3\, |W|^2\Big)  \,,\,
\eea
where the covariant derivatives are defined with respect to all the chiral variables on which the K\"{a}hler potential ($K$) and the holomorphic superpotential ($W$) generically depend on. This general expression had resulted in a series of the so-called ``master-formulae" for the scalar potential for a given set of K\"ahler- and the super-potentials; e.g.~see \cite{Cicoli:2007xp,Shukla:2015hpa,Shukla:2016hyy,Cicoli:2017shd,Gao:2017gxk,Shukla:2019wfo,AbdusSalam:2020ywo,Cicoli:2021dhg,Cicoli:2021tzt,Marchesano:2020uqz,Leontaris:2022rzj,Leontaris:2023lfc}. 
For computing the scalar potential in a given model we need several ingredients which we briefly recollect in this section.

\subsection{Forms, fluxes, and moduli}
The massless states in the four dimensional effective theory are in one-to-one correspondence with harmonic forms which are either  even or odd under the action of an isometric, holomorphic involution ($\sigma$) acting on the internal compactifying CY threefolds ($X$), and these do generate the equivariant  cohomology groups $H^{p,q}_\pm (X)$. For that purpose, let us fix our conventions, and denote the bases  of even/odd two-forms as $(\mu_\alpha, \, \nu_a)$ while four-forms as $(\tilde{\mu}_\alpha, \, \tilde{\nu}_a)$ where $\alpha\in h^{1,1}_+(X), \, a\in h^{1,1}_-(X)$. Also, we denote the zero- and six- even forms as ${\bf 1}$ and $\Phi_6$ respectively. In addition, the bases for the even and odd cohomologies  of three-forms $H^3_\pm(X)$ are denoted as the symplectic pairs $(a_K, b^J)$ and $({\cal A}_\Lambda, {\cal B}^\Delta)$ respectively. Here we fix the normalization in the various cohomology bases as,
\bea
\label{eq:intersection}
& & \int_X \, \mu_\alpha \wedge \tilde{\mu}^\beta = {\delta}_\alpha^{\, \, \, \beta} , \qquad \int_X \, \nu_a \wedge \tilde{\nu}^b = {\delta}_a^{\, \, \,b},\\
& & \int_X \, \mu_\alpha \wedge \mu_\beta \wedge \mu_\gamma = \ell_{\alpha \beta \gamma}, \qquad \int_X \, \mu_\alpha \wedge \nu_a \wedge \nu_b = \hat{\ell}_{\alpha a b}, \quad \int_X \Phi_6 = 1, \, \, \nonumber\\
& & \int_X a_K \wedge b^J = \delta_K{}^J, \qquad \int_X {\cal A}_\Lambda \wedge {\cal B}^\Delta = \delta_\Lambda{}^\Delta. \nonumber
\eea
Here, for the orientifold choice with $O3/O7$-planes, $K\in \{1, ..., h^{2,1}_+\}$ and $\Lambda\in \{0, ..., h^{2,1}_-\}$ while for $O5/O9$-planes, one has $K\in \{0, ..., h^{2,1}_+\}$ and $\Lambda\in \{1, ..., h^{2,1}_-\}$. It has been observed that setups with odd-moduli $G^a$ corresponding to $h^{1,1}_-(X) \neq 0$ are usually less studied as compared to the relatively simpler case of $h^{1,1}_-(X) = 0$, and explicit construction of such CY orientifolds with odd two-cycles can be found in \cite{Lust:2006zg,Lust:2006zh,Blumenhagen:2008zz,Cicoli:2012vw,Gao:2013rra,Gao:2013pra}.

Now, the various field ingredients can be expanded in appropriate bases of the equivariant cohomologies. For example, the K\"{a}hler form $J$, the two-forms $B_2$,  $C_2$ and the RR four-form $C_4$ can be expanded as \cite{Grimm:2004uq}
\bea
\label{eq:fieldExpansions}
& & J = t^\alpha\, \mu_\alpha, \quad  B_2= b^a\, \nu_a , \quad C_2 =c^a\, \nu_a, \quad C_4 = {\rho}_{\alpha} \, \tilde\mu^\alpha + \cdots, 
\eea
where $t^\alpha$, and $\{b^a, \, c^a, \rho_\alpha\}$ denote the Einstein-frame two-cycle volume moduli, and a set of axions descending from their respective form-potentials $\{B_2, C_2, C_4\}$, while dots $\cdots$ encode the information of a dual pair of spacetime one-forms, and two-form dual to the scalar field $\rho_\alpha$ which are not relevant for the current analysis. 
In addition, we consider the choice of involution $\sigma$ to be such that $\sigma^*\Omega_3 = - \Omega_3$, where $\Omega_3$ denotes the nowhere vanishing holomorphic three-form depending on the complex structure moduli $U^{i}$ counted in the $h^{2,1}_-(X)$ cohomology. Using these pieces of information one defines a set ($U^i, S, G^a, T_\alpha$) of the chiral coordinates as below \cite{Benmachiche:2006df},
\bea
\label{eq:N=1_coords}
& & \hskip-2cm U^i = v^i - i\, u^i, \qquad S \equiv C_0 + \, i \, e^{-\phi} = C_0 + i\, s, \qquad G^a= c^a + S \, b^a \, ,\\
& & \hskip-2cm T_\alpha= \left({\rho}_\alpha +  \hat{\ell}_{\alpha a b} c^a b^b + \frac{1}{2} \, S \, \hat{\ell}_{\alpha a b} b^a \, b^b \right)  -\frac{i}{2} \, \ell_{\alpha\beta\gamma} t^\beta t^\gamma\, ,\nonumber
\eea
where the triple intersection numbers $\ell_{\alpha\beta\gamma}$ and $\hat{\ell}_{\alpha a b}$ are defined in Eq.~(\ref{eq:intersection}), and using $\ell_{\alpha\beta\gamma}$, the Einstein-frame overall volume (${\cal V}$) of the internal background can be generically written in terms of the two-cycle volume moduli as below,
\begin{eqnarray}
& & {\cal V} = \frac{1}{6} \,{\ell_{\alpha \beta \gamma} \, t^\alpha\, t^\beta \, t^{\gamma}}.
\end{eqnarray}
 Using appropriate chiral variables ($U^i, S, G^a, T_\alpha$) as defined in (\ref{eq:N=1_coords}), a generic form of the tree-level K\"{a}hler potential can be written as below,
\bea
\label{eq:K}
& & \hskip-2.0cm K = -\ln\left(i\int_{X}\Omega_3\wedge{\bar\Omega_3}\right) - \ln\left(-i(S-\ov S)\right) -2\ln{\cal V} \,.
\eea
Here, the nowhere vanishing involutively-odd holomorphic three-form $\Omega_3$, which generically depends on the complex structure moduli ($U^i$), can be given as below, 
\bea
\label{eq:Omega3}
& &  \Omega_3\, \equiv  {\cal X}^\Lambda \, {\cal A}_\Lambda - \, {\cal F}_{\Lambda} \, {\cal B}^\Lambda \,. 
\eea
Here, the period vectors $({\cal X}^\Lambda, {\cal F}_\Lambda)$ are encoded in a pre-potential (${\cal F}$) of the following form,
\bea
\label{eq:prepotential}
& & \hskip-1cm {\cal F} = ({\cal X}^0)^2 \, \, f({U^i}) \,, \qquad  f({U^i}) = \frac{1}{6}\,{{l}_{ijk} \, U^i\, U^j \, U^k} +  \frac{1}{2} \,{\tilde{p}_{ij} \, U^i\, U^j} +  \,{\tilde{p}_{i} \, U^i} +  \frac{1}{2} \,{i\, \tilde{p}_0} \,.
\eea 
In fact, the function $f({U^i})$ can generically have an infinite series of non-perturbative contributions, which we ignore for the current work assuming to be working in the large complex structure limit. The quantities $\tilde{p}_{ij}, \tilde{p}_i$ and $\tilde{p}_0$ are real numbers where $\tilde{p}_0$ is related to the perturbative $(\alpha^\prime)^3$-corrections on the mirror side \cite{Hosono:1994av,Arends:2014qca, Blumenhagen:2014nba}. Further, the chiral coordinates $U^i$'s are defined as $U^i =\frac{\delta^i_\Lambda \, {\cal X}^\Lambda}{{\cal X}^0}$ where ${l}_{ijk}$'s are triple intersection numbers on the mirror (CY) threefold. With these pieces of information, the K\"ahler potential (\ref{eq:K}) takes the following explicit form in terms of the respective ``saxions" of the chiral variables defined in (\ref{eq:N=1_coords}),
\bea
\label{eq:K-explicit}
& & \hskip-1cm K = -\ln \left(\frac{4}{3} \, l_{ijk}\, u^i u^j u^k + 2\, \tilde{p}_0 \right)\, - \, \ln (2\,s) - 2 \ln\left(\frac{1}{6} \,{\ell_{\alpha \beta \gamma} \, t^\alpha\, t^\beta \, t^{\gamma}}\right).
\eea

\subsection{T-dual fluxes and the superpotential}
In this subsection, first we recollect the relevant features of the minimal type IIB flux superpotential induced by the standard three-form fluxes $(F_3, H_3)$ \cite{Gukov:1999ya,Dasgupta:1999ss}, along with the inclusion of additional fluxes via T-dual completion arguments. Taking the choice of orientifold action resulting in $O3/O7$ type setting, one finds that one can generically have the following non-trivial flux-components \cite{Aldazabal:2006up,Robbins:2007yv,Blumenhagen:2015kja},
\bea
\label{eq:allowedFluxes}
& & \left(F_\Lambda, F^\Lambda\right), \quad \left(H_\Lambda, H^\Lambda\right), \quad \left({\omega}_a{}^\Lambda, {\omega}_{a \Lambda}\right), \quad \left(\hat{Q}^{\alpha{}\Lambda} , \,\, \hat{Q}^{\alpha}{}_{\Lambda}\right), \\
& & \left(\hat{\omega}_\alpha{}^K, \hat{\omega}_{\alpha K}\right), \quad  \left({Q}^{a{}K}, \, {Q}^{a}{}_{K} \right), \quad \left(R_K, R^K \right)\,. \nonumber
\eea
Here, the fluxes in the first line of (\ref{eq:allowedFluxes}) are relevant for the F-term contributions through a holomorphic superpotential while the ones in the second line induce the D-term effects \cite{Robbins:2007yv}. In addition, one can have S-dual completion of this setting via inclusion of the so-called $P$-flux with its non-trivial components being given as: $({P}^{a{}K}, \, {P}^{a}{}_{K}, \, \hat{P}^{\alpha{}\Lambda} , \, \hat{P}^{\alpha}{}_{\Lambda})$ \cite{Aldazabal:2006up,Aldazabal:2008zza,Shukla:2015rua}. Now, focusing on the class of orientifold setups with $h^{2,1}_+(X) =0$\footnote{For models with $h^{2,1}_+(X) \neq 0$ having non-geometric $R$-flux and the possibility of D-term contributions see \cite{Robbins:2007yv, Shukla:2015bca, Shukla:2015rua, Shukla:2015hpa, Blumenhagen:2015lta}.}, the type IIB generalized flux superpotential can be given as below,
\bea
\label{eq:W-Tdual}
& & \hskip-0.5cm W = \biggl[\ov{F}_0 + \, U^i \, \ov{F}_i + \frac{1}{2} \, l_{ijk} U^i U^j\, F^k - \frac{1}{6} \, l_{ijk} U^i U^j U^k\, F^0 - i\, \tilde{p}_0 \, F^0 \biggr] \\
& & - \, S \biggl[\ov{H}_0 + \, U^i \, \ov{H}_i + \frac{1}{2} \, l_{ijk} U^i U^j \, H^k - \frac{1}{6} \, l_{ijk} U^i U^j U^k \, H^0 - i\, \tilde{p}_0 \, H^0 \biggr] \nonumber\\
& & - \, G^a \biggl[\ov{\omega}_{a0} + \, U^i \, \ov{\omega}_{ai} + \frac{1}{2} \, l_{ijk} U^i U^j \,\omega_{a}{}^k - \frac{1}{6} \, l_{ijk} U^i U^j U^k\, \omega_{a}{}^0 - i\, \tilde{p}_0 \, \omega_a{}^0 \biggr] \nonumber\\
& & +  \, T_\alpha \biggl[\ov{{Q}}^\alpha{}_0 + \, U^i \, \ov{{Q}}^\alpha{}_i + \frac{1}{2} \, l_{ijk} U^i U^j {Q}^{\alpha \, k} - \frac{1}{6} l_{ijk} U^i U^j U^k \, {Q}^{\alpha 0} - i\, \tilde{p}_0 \, {Q}^{\alpha 0} \biggr], \nonumber
\eea
where given that the complex structure moduli dependent sector is modified by the $\alpha^\prime$-corrections on the mirror side, one needs to consider a set of rational shifts in some of the usual fluxes in (\ref{eq:W-Tdual}) which are given as \cite{Shukla:2019wfo},
\bea
\label{eq:IIB-W-fluxshift}
& & \ov{F}_0 = F_0 - \tilde{p}_i \, F^i\,, \qquad \qquad \ov{F}_i = F_i - \tilde{p}_{ij}\, F^j - \tilde{p}_i\, F^0\,, \\
& & \ov{H}_0 = H_0 - \tilde{p}_i \, H^i\,, \qquad \quad \, \, \, \ov{H}_i = H_i - \tilde{p}_{ij}\, H^i - \tilde{p}_i H^0\,,\nonumber\\
& & \ov\omega_{a0} = \omega_{a0} - \tilde{p}_i\, \, \omega_a{}^i \,, \qquad \quad  \ov\omega_{ai} = \omega_{ai} - \tilde{p}_{ij}\, \omega_a{}^j - \tilde{p}_i\, \omega_a{}^0\,,\nonumber\\
& & \ov{{Q}}^\alpha{}_0= {Q}^\alpha{}_0 - \tilde{p}_i \, {Q}^{\alpha i} \,, \qquad \, \, \ov{{Q}}^\alpha{}_i = {Q}^\alpha{}_i - \tilde{p}_{ij}\, {Q}^{\alpha j} - \tilde{p}_i \, {Q}^{\alpha 0}\,.\nonumber
\eea
{\it In addition to having $h^{2,1}_+(X) = 0$ to avoid D-term effects, in the current work, we will be interested in orientifolds with trivial $(1,1)$ cohomology in the odd-sector, and therefore our current setup does not include the odd moduli ($G^a$). For the purpose of studying the scalar potential, we will also make another simplification in our superpotential by considering the fluxes to adopt appropriate rational values in order to absorb the respective rational shifts mentioned in (\ref{eq:IIB-W-fluxshift}); for example, see \cite{Blumenhagen:2014nba,Escobar:2018rna} regarding studies without including the non-geometric fluxes.} Subsequently, in the large complex structure limit, we can fairly use the following form of the superpotential,
\bea
\label{eq:W-Tdual-simp}
& & \hskip-0.5cm W = \biggl[{F}_0 + \, U^i \, {F}_i + \frac{1}{2} \, l_{ijk} U^i U^j\, F^k - \frac{1}{6} \, l_{ijk} U^i U^j U^k\, F^0 \biggr] \\
& & - \, S \biggl[{H}_0 + \, U^i \, {H}_i + \frac{1}{2} \, l_{ijk} U^i U^j \, H^k - \frac{1}{6} \, l_{ijk} U^i U^j U^k \, H^0  \biggr] \nonumber\\
& & +  \, T_\alpha \biggl[{{Q}}^\alpha{}_0 + \, U^i \, {{Q}}^\alpha{}_i + \frac{1}{2} \, l_{ijk} U^i U^j {Q}^{\alpha \, k} - \frac{1}{6} l_{ijk} U^i U^j U^k \, {Q}^{\alpha 0} \biggr]\,. \nonumber
\eea
Finally, let us mention that using the dictionary presented in \cite{Shukla:2019wfo} one can equivalently read-off the T-dual completed version of the type IIA flux superpotential, which is a holomorphic function of four types of chiral variables $\{{\rm T}^a, {\rm N}^0, {\rm N}^k, {\rm U}_\lambda\}$ respectively correlated with the set of complexified moduli $\{U^i, S, G^a, T_\alpha\}$ in the type IIB setup. For interested readers we present the T-duality rules for relevant fluxes (appearing in the F-term contributions) in Table \ref{tab_T-dual-fluxes}. 

\noindent
\begin{table}[H]
\begin{center}
\begin{tabular}{|c||c|c|c|c||c|c|c|c|c|c|c|c||} 
\hline
& &&&&&&&&&&&\\
{\rm IIB} & $\, F_0$ & $F_i$ & $\,F^i$ & $F^0$ & $H_0$ & $H_i$ & ${H}^i$ & ${H}^0$ & ${Q}^\alpha{}_0$  & ${Q}^\alpha{}_i$ & ${Q}^{\alpha i}$ & ${Q}^{\alpha 0}$ \\
& &&&&&&&&&&&\\
\hline
& &&&&&&&&&&&\\
{\rm IIA} & $e_0$ & $e_a$ & $m^a$ & $-m^0$ & ${\rm H}_0$ & $w_{a0}$ & ${\rm Q}^{a}{}_0$  & $-{\rm R}_0$ & ${\rm H}^\lambda$ & $w_a{}^\lambda$ & ${\rm Q}^{a \lambda}$  & $-{\rm R}^\lambda$ \\
& &&&&&&&&&&&\\
\hline
\end{tabular}
\end{center}
\caption{A dictionary between the type IIA and type IIB fluxes \cite{Shukla:2019wfo}.}
\label{tab_T-dual-fluxes}
\end{table}


\section{U-dual completion of the flux superpotential}
\label{sec_Udual}
In the previous section we have presented the T-duality transformations among various ingredients of non-geometric type IIA/IIB superpotentials. In this section we extend this analysis with the inclusion of some more fluxes which one needs for establishing the S-duality invariance of the type IIB effective potential. This at the same time demands to include more fluxes on the type IIA side via imposing the T-duality rules on the type IIB side. Let us elaborate more on this point.

The four-dimensional effective potential of the type IIB theory generically have an S-duality invariance following from the underlying ten-dimensional supergravity, and this corresponds to the following $SL(2, \mathbb{Z})$ transformation,
\bea
\label{eq:SL2Za}
& & \hskip-1.5cm S\to \frac{a S+ b}{c S + d}\, \quad \quad {\rm where} \quad a d- b c = 1\,;\quad a,\ b,\ c,\ d\in \mathbb{Z}\,.
\eea
Subsequently it turns out that the complex structure moduli $U^i$'s  and the Einstein-frame volumes (and hence the $T_\alpha$ moduli) are invariant under the $SL(2, \IZ)$ transformation, in the absence of odd-moduli $G^a$ \cite{Grimm:2007xm}. Subsequently, using the transformation,
\bea
& & \left(S - \ov S \right)^{-1} \to |c\, S + d|^2 \, \left(S - \ov S \right)^{-1}\,,
\eea
one finds that the K\"ahler potential (\ref{eq:K}) transforms as:
\bea
\label{eq:Kpot-Sduality}
e^K \longrightarrow |c \, S + d|^2 \, e^K\,.
\eea
Moreover, these $SL(2, \IZ)$ transformations have two generators which can be understood with distinct physical significance as below, 
\bea
& \hskip-3cm {\bf (S1)} \qquad & S \to S +1:  \qquad \qquad {\rm Gen}_1 = \left(\begin{array}{cc} 1  & \quad 1 \\ 0 & \quad 1 \end{array}\right); \\
& \hskip-3cm {\bf (S2)} \qquad & S \to -\frac{1}{S}:  \qquad \qquad \quad {\rm Gen}_2= \left(\begin{array}{cc} 0  & \quad -1 \\ 1 & \quad 0 \end{array}\right). \nonumber
\eea
The first transformation {\bf (S1)} simply corresponds to an axionic shift in the universal axion $C_0$, namely $C_0 \to C_0 + 1$ and it is not of much physical significance. However, the second transformation, which is also known as the strong-weak duality or the S-duality is quite crucial and interesting physical implications. For example, demanding the physical quantities such as the gravitino mass-square ($m_{3/2}^2\sim e^K |W|^2$) to be invariant under S-duality demands the superpotential $W$ to be a holomorphic function with modularity of weight $-1$ which means \cite{Font:1990gx, Cvetic:1991qm, Grimm:2007xm},
\bea
\label{eq:SupW-Sduality}
& & \hskip-2.0cm W \to \frac{W}{c \, S + d}.
\eea
This further implies that the various fluxes possibly appearing in the superpotential have to readjust among themselves to respect this modularity condition (\ref{eq:SupW-Sduality}), and one such S-dual pair of fluxes in the type IIB framework is the so-called $(F, H)$ consisting of the RR and NS-NS three-form fluxes transforming in the following manner,
\bea
& & \hskip-1.5cm \left(\begin{array}{c} F \\ H \end{array}\right) \longrightarrow \left(\begin{array}{cc} a  & \quad b \\ c & \quad d \end{array}\right)
\left(\begin{array}{c} F \\ H \end{array}\right);\, \qquad {\bf (S2)} \Longrightarrow \{F \to - \, H, \quad H \to F\}. 
\eea
In fact it turns out that making successive applications of T/S-dualities results in the need of introducing more and more fluxes compatible with (\ref{eq:SupW-Sduality}) such that the superpotential not only receives cubic couplings for the $U^i$-moduli but also for the $T_\alpha$-moduli \cite{Aldazabal:2006up}. In fact, it turns out that one needs a total of four S-dual pairs of fluxes, commonly denotes as:  $(F, H), \, (Q, P), \, (P^\prime, Q^\prime)$ and $(H^\prime, F^\prime)$ \cite{Aldazabal:2006up,Aldazabal:2008zza,Aldazabal:2010ef,Lombardo:2016swq,Lombardo:2017yme,Leontaris:2023lfc}. In this section we will elaborate more on it in some detail. 

\subsection{Insights from the non-symplectic (toroidal) formulation}

Flux superpotentials with the U-dual completion \cite{Aldazabal:2006up,Aldazabal:2008zza} have been studied at various occasions, mostly in the framework of toroidal constructions \cite{Aldazabal:2010ef,Lombardo:2016swq,Lombardo:2017yme,Leontaris:2023lfc} . Using the standard flux formulation in which fluxes are expressed in terms of the real six-dimensional indices, one can denote the four pairs of S-dual fluxes with the following index structure,
\bea
\label{eq:All-flux}
& & F_{ijk}, \qquad H_{ijk}, \qquad Q_i{}^{jk}, \qquad P_i{}^{jk}, \\
& & P'^{i,jklm}, \qquad Q'^{i,jklm}, \qquad  H'^{ijk,lmnpqr}, \qquad  F'^{ijk,lmnpqr},\nonumber
\eea
and therefore, one can consider $(P', Q')$ fluxes as some $(1,4)$ mixed-tensors in which only the last four-indices are anti-symmetrized, while $(H', F')$ flux can be understood as some $(3, 6)$ mixed-tensors where first three indices and last six indices are separately anti-symmetrized. Further details about the mixed-tensor fluxes can be found in \cite{Lombardo:2016swq,Lombardo:2017yme}.

Subsequently, using generalized geometry motivated through toroidal constructions, it has been argued that the type IIB superpotential governing the dynamics of the four-dimensional effective theory (which respects the invariance under $SL(2, {\mathbb Z})^7$ symmetry) can be given as \cite{Aldazabal:2006up,Aldazabal:2008zza,Aldazabal:2010ef,Lombardo:2016swq,Lombardo:2017yme,Leontaris:2023lfc},
\bea
\label{eq:W-toroidal}
& & W = \int_{X} \, \left(f_+ \, - \, S\, f_- \right) \, \cdot e^{{\cal J}} \wedge \Omega_3 \,,
\eea
where ${\cal J}$ denotes the complexified 4-form ${\cal J} = C_4 - \frac{i}{2} \, J \wedge J \equiv \tilde\mu^\alpha T_\alpha$, and the various flux actions are encoded in the followings quantities $f_\pm$,
\bea
\label{eq:WIIBGENnew2b}
& & f_+  \cdot e^{{\cal J}} 
= F + Q \triangleright {\cal J} +  P^\prime \diamond {\cal J}^2 + H^\prime \odot {\cal J}^3 \,, \\
& & f_- \cdot e^{{\cal J}_c} 
= H + P \triangleright {\cal J} +  Q^\prime \diamond {\cal J}^2 + F^\prime \odot {\cal J}^3 \,.\nonumber
\eea
The explicit form of these flux-actions are elaborated as below,
\bea
\label{eq:fluxactionsIIB-udual-old}
&& \left(Q\triangleright {\cal J} \right)_{a_1a_2a_3} = \frac{3}{2} \, Q^{b_1b_2}_{[\underline{a_1}} \, {\cal J}_{\underline{a_2} \, \underline{a_3}] b_1 b_2} \, ,\\
&& \left(P^\prime \diamond {\cal J}^2 \right)_{a_1a_2a_3} = \frac{1}{4} \, {P^\prime}^{c,b_1b_2b_3b_4}  \, {\cal J}_{[\underline{a_1} \underline{a_2} |c b_1|}\, {\cal J}_{\underline{a_3}] b_2 b_3 b_4}\,, \nonumber\\
&& \left(H^\prime \odot {\cal J}^3 \right)_{a_1a_2a_3} = \frac{1}{192} \, {H^\prime}^{c_1 c_2 c_3, b_1 b_2b_3b_4b_5b_6}  \, {\cal J}_{[\underline{a_1}\underline{a_2} | c_1 c_2|}\, {\cal J}_{\underline{a_3}]c_3b_1b_2} \, {\cal J}_{b_3 b_4 b_5 b_6}\,, \nonumber
\eea
and the remaining flux actions $\left(P\triangleright {\cal J} \right)$, $\left(Q^\prime \diamond {\cal J}^2\right)$ and $\left(F^\prime \odot {\cal J}^3\right)$ are defined similarly as to the flux actions for $\left(Q\triangleright {\cal J}\right)$, $\left(P^\prime \diamond {\cal J}^2\right)$ and $\left(H^\prime \odot {\cal J}^3\right)$ respectively. Let us mention that now our first task is to understand/rewrite the flux actions (\ref{eq:fluxactionsIIB-udual-old}) in terms of symplectic ingredients. In this regard, we mention the following useful identities which have been utilized in understanding the connection between the Heterotic superpotential and the type IIB superpotenial with the U-dual fluxes in \cite{Aldazabal:2010ef}:
\bea
\label{eq:Jidentity}
& & J_{p_1 p_2} = \frac{1}{4^2. \, 4 !} \, J^2_{i_1 i_2 i_3 i_4} \, J^2_{i_5 i_6 p_1 p_2} \, {\cal E}^{i_1 i_2 i_3 i_4 i_5 i_6} \, ,\\
& & \hskip-1.18cm J^3_{p_1 p_2 p_3 p_4 p_5 p_6} = \frac{5}{128} \, J^2_{i_1 i_2 i_3 i_4} \, J^2_{i_5 i_6 [\underline{p_1 p_2}} \, J^2_{\underline{p_3 p_4 p_5 p_6}]} \, {\cal E}^{i_1 i_2 i_3 i_4 i_5 i_6} \,.\nonumber
\eea
As argued in \cite{Aldazabal:2010ef}, these identities are generically true, even for the beyond toroidal cases as well. Moreover, simple volume scaling arguments suggest that ${\cal E}^{i_1 i_2 i_3 i_4 i_5 i_6}$ is a volume dependent quantity satisfying the following useful identity,
\bea
& & {\cal E}^{i_1 i_2 i_3 i_4 i_5 i_6} = \frac{\epsilon^{i_1 i_2 i_3 i_4 i_5 i_6}}{\sqrt{|{\rm det}\, g|}} = \frac{\epsilon^{i_1 i_2 i_3 i_4 i_5 i_6}}{{\cal V}} \,,
\eea
where $\epsilon^{i_1 i_2 i_3 i_4 i_5 i_6}$ denotes the six-dimensional anti-symmetric Levi-Civita symbol. This normalisation by a volume factor can also be understood through the following relation satisfied by the antisymmetric Levi-Civita symbol $\epsilon^{ijklmn}$ and the internal metric,
\bea
\label{eq:Identity-1}
& & \hskip-1cm \epsilon^{ijklmn} g_{i i'} g_{j j'} g_{k k'} g_{l l'} g_{m m'} g_{n n'} = |{\rm det}\, g|\, \epsilon_{i'j'k'l'm'n'} =  {\cal V}^2\, \epsilon_{i'j'k'l'm'n'},
\eea
which is equivalent to,
\bea
& & \hskip-1cm {\cal E}^{ijklmn} g_{i i'} g_{j j'} g_{k k'} g_{l l'} g_{m m'} g_{n n'} = {\cal E}_{i'j'k'l'm'n'}.
\eea
Further, it has been observed from the toroidal results about studying the taxonomy of the various scalar potential pieces in \cite{Leontaris:2023lfc} that the prime fluxes can be equivalently expressed in another way using the Levi-Civita tensor given as below\footnote{As opposed to using Levi-Civita symbol $\epsilon^{ijklmn}$ in \cite{Leontaris:2023lfc}, here we use the Levi-Civita tensor ${\cal E}^{ijklmn}$ in defining the fluxes in (\ref{eq:shortPrimed-flux}). The reason will be more clear when we discuss the cohomology formulation of these fluxes later on.},
\bea
\label{eq:shortPrimed-flux}
& & P'_{ij}{}^k = \frac{1}{4!} \,  {\cal E}_{ijlmnp}\, P'^{k,lmnp}, \qquad Q'_{ij}{}^k = \frac{1}{4!} \,   {\cal E}_{ijlmnp}\, Q'^{k,lmnp}, \\
& & H'^{ijk} = \frac{1}{6!} \,   {\cal E}_{lmnpqr}\, H'^{ijk,lmnpqr}, \qquad F'^{ijk} = \frac{1}{6!} \,   {\cal E}_{lmnpqr}\, F'^{ijk,lmnpqr}. \nonumber
\eea
The first thing to observe about these redefinitions is the fact that the index structure of $(P'_{ij}{}^k, Q'_{ij}{}^k)$ looks similar to those of the so-called geometric fluxes ($\omega_{ij}{}^k$) while the remaining prime fluxes $(H'^{ijk}, F'^{ijk})$ have the index structure similar to those of the non-geometric $R$-fluxes as motivated in Eq.~(\ref{eq:Tdual}) following from the chain of successive T-dualities applied to the three-form $H$ flux. Note that, the presence of ${\cal E}_{ijlmnp}$ introduces a volume dependence in the redefined version of the prime fluxes, which helps in taking care of the overall volume factor appearing repeatedly in the following up equations of the various scalar potential pieces via producing a common overall factor depending on volume for all the pieces. This subsequently results in having an overall factor of ${\cal V}^{-2}$ for all the topological pieces, and a factor of ${\cal V}^{-1}$ for the remaining (non-topological) pieces as seen in the non-symplectic formulation in \cite{Leontaris:2023lfc}.  However, while expressing the superpotential (which is a holomorphic function of the chiral variables) using such volume dependent fluxes $P'_{ij}{}^k$ etc.~as defined in (\ref{eq:shortPrimed-flux}), one has to be a bit careful and appropriately take care of the volume dependent factor. On these lines, it might be worth mentioning that \cite{Aldazabal:2010ef} uses the same symbol ``$\, \epsilon^{ijklmn}\, $" for the identities given in (\ref{eq:Jidentity}) as well as for defining the prime flux actions similar to the ones we consider in (\ref{eq:fluxactionsIIB-udual-old}).This indicates that the prime fluxes defined in \cite{Aldazabal:2010ef} can have an overall volume factor (at least in the toroidal case) in one of the two formulations, and the holomorphicity of the superpotential has to be respected via appropriately taking care of the presence of the overall volume (${\cal V}$) factors. On these lines, it is worth mentioning that the prime fluxes of the form (\ref{eq:All-flux}), i.e. without the overall volume factors, are considered in \cite{Lombardo:2016swq,Lombardo:2017yme} and this formulation does not need any extra volume factor to keep the superpotential holomorphic. 

Here let us also note the fact that the identities presented in Eq.~(\ref{eq:Jidentity}) are expressed in terms of the real six-dimensional indices, and we need a cohomology/symplectic version of these identities as well as the new fluxes defined in (\ref{eq:shortPrimed-flux}), similar to what we have argued for the flux actions defined in (\ref{eq:fluxactionsIIB-udual-old}). 

\subsection{Symplectic formulation of fluxes and the superpotential}
Having learnt the lessons from the toroidal setup, now we briefly discuss the U-dual completion of the flux superpotential via taking a symplectic approach. 

\subsubsection*{Step-0}
To begin with we consider the standard GVW flux superpotential generated by the S-dual pair of $(F, H)$ fluxes given as below \cite{Gukov:1999ya},
\bea
\label{eq:WFH}
& & \hskip-1.0cm W_0   = \int_{X} \biggl[\left({F} - S \, {H}\right) \biggr]_3 \wedge \Omega_3\,.
\eea
This results in the so-called ``no-scale structure" in the scalar potential which receives a dependence on the overall volume of the internal background only via $e^K$ factor, and hence scales as ${\cal V}^{-2}$. There is no superpotential coupling for the K\"ahler moduli which remain flat in the presence of $(F, H)$ fluxes in the internal background.

\subsubsection*{Step-1}
In order to break the no-scale structure and induced volume moduli dependence pieces in the scalar potential, one subsequently includes the non-geometric $Q$-fluxes. In the absence of odd-moduli, the type IIB non-geometric flux superpotential takes the following form \cite{Aldazabal:2006up},
\bea
\label{eq:WFHQ}
& & \hskip-1.0cm W_1   = \int_{X} \biggl[\left({F} - S \, {H} \right) + \, {Q}^{\alpha}  \,{T}_\alpha \biggr]_3 \wedge \Omega_3\,,
\eea
where the quantities in the bracket $\bigl[...\bigr]_3$ are three-forms which can be expanded in an appropriate basis as below,
\bea
\label{eq:FluxAction-1}
&& \left(Q\triangleright {\cal J} \right) = {Q}^{\alpha}\, T_\alpha, \qquad \qquad {Q}^{\alpha} = -\, {Q}^{\alpha{}\Lambda} \, {\cal A}_\Lambda + {Q}^{\alpha}{}_{\Lambda} \, {\cal B}^\Lambda\,.
\eea
Recall that the various $Q$-flux components surviving under the orientifold-action can be given as: $Q\equiv \left({Q}^{\alpha{}\Lambda} , \, {Q}^{\alpha}{}_{\Lambda}\right)$ as the odd sector of $(1,1)$-cohomology is trivial. The expanded version of this T-dual completed superpotential (\ref{eq:WFHQ}) is already presented in (\ref{eq:W-Tdual-simp}), and its type IIA analogue can be obtained by simply using the dictionary given in Table \ref{tab_T-dual-fluxes}, along with the T-duality rules among the chiral variables defined as: $S \leftrightarrow {\rm N}^0, \, U^i  \leftrightarrow {\rm T}^a$ and $T_\alpha \leftrightarrow {\rm U}_\lambda$. 

Now, we further take the iterative steps of T-and S-dualities to reach the U-dual completion of the type IIB superpotential.

\subsubsection*{Step-2}
Note that the GVW flux superpotential (\ref{eq:WFH}) respects the underlying S-duality in the type IIB description, however the inclusion of non-geometric $Q$-flux, which leads to the flux superpotential (\ref{eq:WFHQ}), does not retain the S-duality invariance of the theory. For that purpose, the simplest S-dual completion of the flux superpotential (\ref{eq:WFHQ}) can be given by adding a new set of non-geometric flux, namely the so-called $P$-flux which is S-dual to the NS-NS $Q$-flux. Therefore one has another S-dual pair of fluxes, namely $(Q, P)$ which is similar to the standard $(F, \, H)$ flux-pair, and transforms under the $SL(2, \IZ)$ transformation in the following manner \cite{Aldazabal:2006up,Aldazabal:2008zza, Guarino:2008ik,Blumenhagen:2015kja},
\bea
& & \hskip-1.5cm \left(\begin{array}{c} Q \\ P \end{array}\right) \longrightarrow \left(\begin{array}{cc} a  & \quad b \\ c & \quad d \end{array}\right)
\left(\begin{array}{c} Q \\ P \end{array}\right)\,,  \qquad ad-bc =1.
\eea
Being S-dual to the non-geometric $Q$-flux, such $P$-fluxes have the flux actions similar to those of the $Q$-flux as defined in (\ref{eq:FluxAction-1}). Subsequently a superpotential of the following form is generated,
\bea
\label{eq:WFHQP}
& & \hskip-1.0cm W_2   = \int_{X} \biggl[\left({F} - S \, {H} \right) 
+ \,\left( {Q}^{\alpha} - S  {P}^{\alpha}\right)  \,{T}_\alpha 
 \biggr]_3 \wedge \Omega_3. 
\eea
Using the explicit expressions for the holomorphic three-form ($\Omega_3$) as given Eq.~(\ref{eq:Omega3}), this flux superpotential $W_2$ can be equivalently written in the following form,
\begin{eqnarray}
\label{eq:W2}
& & \hskip-1.0cm W_2  
 =\, \biggl[F_0 + F_i\, U^i + F^i \,\left(\frac{1}{2} \, l_{ijk} U^j U^k\right) - F^0 \, \left(\frac{1}{6} \, l_{ijk} U^i U^j U^k\right) \biggr]\\
& & \, - \, S \, \biggl[H_0 + H_i\, U^i + H^i \, \left(\frac{1}{2} \, l_{ijk} U^j U^k\right) - H^0 \, \left(\frac{1}{6} \, l_{ijk} U^i U^j U^k\right) \biggr] \nonumber\\
& & \, + \, T_\alpha\, \biggl[{Q}^\alpha{}_0 + {Q}^\alpha{}_i\, U^i + {Q}^{\alpha \, i} \, \left(\frac{1}{2} \, l_{ijk} U^j U^k\right) - {Q}^{\alpha 0} \, \left(\frac{1}{6} l_{ijk} U^i U^j U^k\right) \biggr]\,\nonumber\\
& & \, - \, S \, T_\alpha\, \biggl[{P}^\alpha{}_0 + {P}^\alpha{}_i\, U^i + {P}^{\alpha \, i} \, \, \frac{1}{2} l_{ijk} U^j U^k - {P}^{\alpha 0} \, \left(\frac{1}{6} l_{ijk} U^i U^j U^k\right) \biggr]\,. \nonumber
\end{eqnarray} 
The scalar potential induced from this flux superpotential has been studied in \cite{Gao:2015nra,Shukla:2016hyy}.

\subsubsection*{Step-3}
From the T-duality transformations, we know that a piece with moduli dependence of the kind $(S \, T_\alpha)$ on the type IIB side, as we have in Eq.~(\ref{eq:WFHQP}), corresponds to a piece of the kind $({\rm N}^0\, {\rm U}_\lambda)$ on the type IIA side\footnote{We establish the correlation between the type IIA and type IIB superpotentials via considering the T-duality rules for the moduli as $\{{\rm T}^a, {\rm N}^0, {\rm N}^k, {\rm U}_\lambda\} \to \{U^i, S, G^a, T_\alpha\}$, and for the fluxes as in Table \ref{tab_T-dual-fluxes}.}, where such a term can be generated via a quadric in $\Omega_c$ which is linear in ${\rm N}^0$ and ${\rm U}_\lambda$. Here, we recall that $\Omega_c$ is defined by complexifying RR three-from (${\rm C}_3$) axion with the holomorphic CY three-from $\Omega_3$, leading to $\Omega_c = {\rm N}^{\hat k} {\cal A}_{\hat k} - {\rm U}_\lambda {\cal B}^\lambda$ in type IIA setup, e.g. see \cite{Shukla:2019wfo} for more details. But a quadric $\Omega_c^2$ will also introduce a piece on the type IIA side which is quadratic in ${\rm U}$-moduli leading to a quadratic in $T$-moduli on type IIB side, and hence will also introduce some new fluxes on type IIA side which are not T-dual to any of the fluxes ($F, H, Q$ and $P$) introduced so far on the type IIB side. Such a quadratic term in $T$-moduli on the type IIB side, can be of the following kind: 
\bea
& & \int_{X} \biggl[\frac{1}{2} \, {P}^{\prime\alpha\beta} \, \cdot \,{T}_\alpha \, {T}_\beta \biggr]_3 \wedge \Omega_3 \,,
\eea
which results in introducing a new type of flux that we denote as $P^\prime$ flux. Also, for the moment we consider $P^\prime$-flux to be of the form $ {P}^{\prime\alpha\beta}$, just to have proper contractions with $T$-moduli indices. We will discuss some more insights of such $P^\prime$-fluxes while we compare the results with those of the non-symplectic (toroidal) proposal in \cite{Aldazabal:2006up, Aldazabal:2008zza, Aldazabal:2010ef, Lombardo:2016swq, Lombardo:2017yme}. We will follow the same logic for introducing other prime fluxes as we discuss now.

After introducing the $P^\prime$-flux and subsequently demanding the S-duality invariance in type IIB side, we need to introduce the so-called $Q^\prime$-flux which is S-dual of the $P^\prime$-flux, and the hence they form another S-dual pair $(P^\prime, \, Q^\prime)$ which leads to the following term in the flux superpotential,
\bea
& & \int_{X} \biggl[\left({P}^{\prime\alpha\beta} - \, S \, \, {Q}^{\prime\alpha\beta}\right) \cdot \left(\frac{1}{2} \,{T}_\beta \, {T}_\gamma \right) \biggr]_3 \wedge \Omega_3 \,.
\eea
So now, we have a superpotential piece on the type IIB side which has a factor of moduli $(S\, T_\beta \, T_\gamma)$. Subsequently, this corresponds to a type IIA superpotential piece with a moduli factor $({\rm N}^0\, {\rm U}_\lambda \, {\rm U}_\rho)$, and hence is expected to arise from a cubic in $\Omega_c$. However, a cubic in $\Omega_c$ will not only generate this piece but will also additionally generate a piece with moduli factor $({\rm U}_\lambda\, {\rm U}_\rho \, {\rm U}_\gamma)$ in type IIA. Again getting back to the type IIB side, will generate a term with moduli factor of the type $(T_\alpha\, T_\beta \, T_\gamma)$. This will subsequently result in introducing a new type of fluxes, the so-called NS$^\prime$-flux denoted as $H^\prime$, and then completing the S-dual pair via introducing another new flux, namely $F^\prime$-flux, leads to the following superpotential terms,
\bea
& & \int_{X} \biggl[\left({ H}^{\prime\alpha \beta \gamma} - \, S \, \, { F}^{\prime\alpha \beta \gamma}\right) \cdot \left(\frac{1}{6} \,T_{\alpha} \, {T}_\beta \, {T}_\gamma \right) \biggr]_3 \wedge \Omega_3 \,.
\eea
However, let us also note that having a type IIB term with a moduli factor $(S \, T_\alpha \, T_\beta\,  T_\gamma)$ implies that on the type IIA side, one would need to introduce a T-dual term with a moduli factor $({\rm N}^0\, {\rm U}_\lambda \, {\rm U}_\rho\, {\rm U}_\gamma)$ which can be introduced via a quartic in $\Omega_c$, and hence in addition one would need to introduce another set of RR$^\prime$ fluxes, namely $F^\prime_{\rm RR}$-flux on type IIA side. 

In this way we observe that the logic of iteration continues when we demand the S/T-dualities back and forth until we arrive at cubic superpotential couplings in $T$- and $U$-moduli on both the (type IIB and type IIA) sides. On the lines of aforementioned U-dual completions, some detailed studies have been made in \cite{Aldazabal:2006up, Aldazabal:2008zza, Aldazabal:2010ef, Lombardo:2016swq, Lombardo:2017yme, Leontaris:2023lfc}, and here we plan to present a symplectic formulation of the four-dimensional scalar potential. Unlike the toroidal proposal \cite{Leontaris:2023lfc}, this symplectic formulation can be easily promoted/conjectured for the beyond toroidal constructions, for example, in case of the non-geometric CY orientifolds.

To summarize, we need to introduce four pairs of S-dual fluxes, i.e. a set of eight types of fluxes transforming in the following manner under the $SL(2,{\mathbb Z})$ transformations,
\bea
& & \hskip-0.1cm \left(\begin{array}{c} F \\ H \end{array}\right) \longrightarrow \left(\begin{array}{cc} a  & \quad b \\ c & \quad d \end{array}\right)
\left(\begin{array}{c} F \\ H \end{array}\right)\,, \qquad \left(\begin{array}{c} Q \\ P \end{array}\right) \longrightarrow \left(\begin{array}{cc} a  & \quad b \\ c & \quad d \end{array}\right)
\left(\begin{array}{c} Q \\ P \end{array}\right)\,, \\
& & \hskip-0.1cm \left(\begin{array}{c} H^\prime \\ F^\prime \end{array}\right) \longrightarrow \left(\begin{array}{cc} a  & \quad b \\ c & \quad d \end{array}\right)
\left(\begin{array}{c} H^\prime \\ F^\prime \end{array}\right)\,, \qquad \left(\begin{array}{c} P^\prime \\ Q^\prime \end{array}\right) \longrightarrow \left(\begin{array}{cc} a  & \quad b \\ c & \quad d \end{array}\right) \left(\begin{array}{c} P^\prime \\ Q^\prime \end{array}\right)\,, \quad ad - bc = 1.\nonumber
\eea
This leads to the following generalized flux superpotential,
\bea
\label{eq:W-all-1}
& & \hskip-1.0cm W_3   = \int_{X} \Biggl[\left({F}  + {Q}^{\alpha}\, T_\alpha + \frac{1}{2} {P}^{\prime\alpha\beta}\, T_\alpha\, T_\beta \, +  \, \frac{1}{6} H^{\prime\alpha\beta\gamma}\, T_\alpha \, T_\beta \, T_\gamma \right) \\
& & \hskip1cm - \, S \, \left({H}  + {P}^{\alpha}\, T_\alpha +  \frac{1}{2} {Q}^{\alpha\beta}\, T_\alpha\, T_\beta \, +\, \frac{1}{6} \,F^{\prime\alpha\beta\gamma}\, T_\alpha \, T_\beta \, T_\gamma \right) \Biggr]_3 \wedge \Omega_3 \,,\nonumber
\eea
where all the terms appearing inside the bracket $[...]_3$ denote a collection of three-forms to be expanded in the symplectic basis $\{{\cal A}_\Lambda, {\cal B}^\Delta\}$ in the following manner,
\bea
\label{eq:FluxActions-2}
&& {Q}^{\alpha} = -\, {Q}^{\alpha{}\Lambda} \, {\cal A}_\Lambda + {Q}^{\alpha}{}_{\Lambda} \, {\cal B}^\Lambda\,,\qquad \qquad \qquad {P}^{\alpha} = -\, {P}^{\alpha{}\Lambda} \, {\cal A}_\Lambda + {P}^{\alpha}{}_{\Lambda} \, {\cal B}^\Lambda\,,\\
&& {P}^{\prime\beta\gamma}  = -\, {P}^{\prime\beta\gamma{}\Lambda} \, {\cal A}_\Lambda + {P}^{\prime\beta\gamma}{}_{\Lambda} \, {\cal B}^\Lambda\,, \qquad \qquad {Q}^{\prime\beta\gamma}\,= -\, {Q}^{\prime\beta\gamma{}\Lambda} \, {\cal A}_\Lambda + {Q}^{\prime\beta\gamma}{}_{\Lambda} \, {\cal B}^\Lambda\,, \nonumber\\
&& H^{\prime\alpha\beta\gamma} = -\, {H}^{\prime\alpha\beta\gamma{}\Lambda} \, {\cal A}_\Lambda + {H}^{\prime\alpha\beta\gamma}{}_{\Lambda} \, {\cal B}^\Lambda\, , \qquad F^{\prime\alpha\beta\gamma}\, = -\, {F}^{\prime\alpha\beta\gamma{}\Lambda} \, {\cal A}_\Lambda + {F}^{\prime\alpha\beta\gamma}{}_{\Lambda} \, {\cal B}^\Lambda\,. \nonumber
\eea
Now we need to determine the explicit expressions of the various flux actions, especially for the $\{P^\prime, Q^\prime\}$ and $\{H^\prime, F^\prime \}$ on the forms ${\cal J}^2$ and ${\cal J}^3$ respectively. Before we do that let us rewrite the superpotential (\ref{eq:W-all-1}) as below,
\bea
\label{eq:W-all-2}
& & \hskip-1.0cm W_3 = \Biggl[\left({F}_\Lambda  + {Q}^{\alpha}{}_\Lambda\, T_\alpha + \frac{1}{2} {P}^{\prime\alpha\beta}{}_\Lambda\, T_\alpha\, T_\beta \, +  \, \frac{1}{6} H^{\prime\alpha\beta\gamma}{}_\Lambda\, T_\alpha \, T_\beta \, T_\gamma \right)\, {\cal X}^\Lambda \\
& & \hskip1cm + \left({F}^\Lambda  + {Q}^{\alpha\Lambda}\, T_\alpha + \frac{1}{2} {P}^{\prime\alpha\beta\Lambda}\, T_\alpha\, T_\beta \, +  \, \frac{1}{6} H^{\prime\alpha\beta\gamma\Lambda}\, T_\alpha \, T_\beta \, T_\gamma \right)\, {\cal F}_\Lambda \biggr]\nonumber\\
& & -\, S\, \Biggl[\left({H}_\Lambda  + {P}^{\alpha}{}_\Lambda\, T_\alpha + \frac{1}{2} {Q}^{\prime\alpha\beta}{}_\Lambda\, T_\alpha\, T_\beta \, + \, \frac{1}{6} F^{\prime\alpha\beta\gamma}{}_\Lambda\, T_\alpha \, T_\beta \, T_\gamma \right)\, {\cal X}^\Lambda \nonumber\\
& & \hskip1cm + \left({H}^\Lambda  + {P}^{\alpha\Lambda}\, T_\alpha + \frac{1}{2} {Q}^{\prime\alpha\beta\Lambda}\, T_\alpha\, T_\beta \, +  \, \frac{1}{6} F^{\prime\alpha\beta\gamma\Lambda}\, T_\alpha \, T_\beta \, T_\gamma \right)\, {\cal F}_\Lambda \Biggr] \,. \nonumber
\eea
Now this form of the superpotential is linear in the axio-dilaton modulus $S$ and has cubic dependence in moduli $U^i$ and $T_\alpha$ both. To be more precise, the explicit form of the superpotential in terms of all these moduli can be given as below,
\bea
\label{eq:W-all-3}
& & \hskip-0.05cm W_3 = \biggl[F_0 + F_i\, U^i + F^i \,\left(\frac{1}{2} \, l_{ijk} U^j U^k\right) - F^0 \, \left(\frac{1}{6} \, l_{ijk} U^i U^j U^k\right) \biggr]\\
& & \hskip0.5cm - \, S \, \biggl[H_0 + H_i\, U^i + H^i \, \left(\frac{1}{2} \, l_{ijk} U^j U^k\right) - H^0 \, \left(\frac{1}{6} \, l_{ijk} U^i U^j U^k\right) \biggr] \nonumber\\
& & \hskip0.15cm +\, T_\alpha\, \biggl[{Q}^\alpha{}_0 + {Q}^\alpha{}_i\, U^i + {Q}^{\alpha \, i} \, \left(\frac{1}{2} \, l_{ijk} U^j U^k\right) - {Q}^{\alpha 0} \, \left(\frac{1}{6} l_{ijk} U^i U^j U^k\right) \biggr]\,\nonumber\\
& & - \, S \, T_\alpha\, \biggl[{P}^\alpha{}_0 + {P}^\alpha{}_i\, U^i + {P}^{\alpha \, i} \, \, \frac{1}{2} l_{ijk} U^j U^k + {P}^{\alpha 0} \, \left(-\frac{1}{6} l_{ijk} U^i U^j U^k\right) \biggr]\, \nonumber\\
& & + \frac{1}{2}\,T_\alpha T_\beta \biggl[{P}^{\prime \alpha\beta}{}_{0} + {P}^{\prime \alpha\beta}{}_{i}\, U^i + {P}^{\prime \alpha\beta i} \left(\frac{1}{2} \, l_{ijk} U^j U^k \right) - {P}^{\prime \alpha\beta 0} \, \left(\frac{1}{6} l_{ijk} U^i U^j U^k\right) \biggr]\, \nonumber\\
& & - \frac{S}{2}\, T_\alpha T_\beta \biggl[{Q}^{\prime \alpha\beta}{}_{0} + {Q}^{\prime \alpha\beta}{}_{i}\, U^i + {Q}^{\prime \alpha\beta i} \left(\frac{1}{2} \, l_{ijk} U^j U^k \right) -{Q}^{\prime \alpha\beta 0} \, \left(\frac{1}{6} l_{ijk} U^i U^j U^k\right) \biggr]\, \nonumber\\
& & + \frac{1}{6} \,T_\alpha\, T_\beta \, T_\gamma \biggl[{H}^{\prime\alpha \beta \gamma}{}_0 + {H}^{\prime\alpha \beta \gamma}{}_i\, U^i + {H}^{{\prime\alpha \beta \gamma} \, i} \left(\frac{1}{2} \, l_{ijk} U^j U^k \right) - {H}^{{\prime\alpha \beta \gamma} 0} \, \left(\frac{1}{6} l_{ijk} U^i U^j U^k\right) \biggr] \nonumber\\
& & - \frac{S}{6}\, T_\alpha\, T_\beta \, T_\gamma \biggl[{F}^{\prime\alpha \beta \gamma}{}_0 + {F}^{\prime\alpha \beta \gamma}{}_i\, U^i + {F}^{{\prime\alpha \beta \gamma} \, i} \left(\frac{1}{2} \, l_{ijk} U^j U^k \right) - {F}^{{\prime\alpha \beta \gamma} 0} \, \left(\frac{1}{6} l_{ijk} U^i U^j U^k\right) \biggr] \,. \nonumber
\eea
Now let us note that the superpotential given in Eq.~(\ref{eq:W-all-1}) can be also rewritten in the following compact form,
\bea
\label{eq:W-all-4}
& & W = \int_{X} \, \biggl[\left(F - S\, H \right) + \left(Q - S\, P \right)\triangleright {\cal J} + \left(P^\prime - S\, Q^\prime \right)\diamond {\cal J}^2 \nonumber\\
& & \hskip2.5cm + \left(H^\prime - S\, F^\prime \right)\odot {\cal J}^3\biggr] \wedge \Omega_3\,,
\eea
where we propose the symplectic form of the various flux actions to be defined as below,
\bea
\label{eq:FluxActions-3}
&& \left(Q\triangleright {\cal J} \right) = T_\alpha \, {Q}^{\alpha}, \qquad \qquad \qquad \qquad \, \, \left(P\triangleright {\cal J} \right) = T_\alpha \, {P}^{\alpha} \, ,\\
&& \left(P^\prime \diamond {\cal J}^2 \right)  = \frac{1}{2} \, {P}^{\prime\beta\gamma}\, T_\beta \, T_\gamma \,, \qquad  \qquad \quad \left(Q^\prime \diamond {\cal J}^2 \right)  = \frac{1}{2} \, {Q}^{\prime\beta\gamma}\, T_\beta \, T_\gamma\,, \nonumber\\
&& \left(H^\prime \odot {\cal J}^3 \right) = \frac{1}{3 !} \, {H}^{\prime\alpha\beta\gamma}\,T_\alpha \, T_\beta \, T_\gamma\, , \qquad \, \, \, \left(F^\prime \odot {\cal J}^3 \right) = \frac{1}{3 !} \, {F}^{\prime\alpha\beta\gamma}\,T_\alpha \, T_\beta \, T_\gamma\,, \nonumber
\eea
and ${Q}^{\alpha}, {P}^{\alpha}, {P}^{\prime\beta\gamma}, {Q}^{\prime\beta\gamma}, {H}^{\prime\alpha\beta\gamma}$ and ${F}^{\prime\alpha\beta\gamma}$ denote the 3-forms as expanded in Eq.~(\ref{eq:FluxActions-2}).

\subsubsection*{More insights of the flux components in the cohomology basis}
Recall that so far in determining the superpotential (\ref{eq:W-all-3}) or equivalently its compact version defined in (\ref{eq:W-all-4})-(\ref{eq:FluxActions-3}), we have only assumed the T-duality rules (among the chiral variables on the type IIB and type IIA side) along with some suitable contractions of $h^{1,1}_+$ indices. In order to explicitly determine the structures of the $P^\prime, Q^\prime, H^\prime$ and $F^\prime$ flux components, we compare our results with the non-symplectic formulation presented in \cite{Aldazabal:2006up, Aldazabal:2008zza, Aldazabal:2010ef, Lombardo:2016swq, Lombardo:2017yme, Leontaris:2023lfc}. In this regard, as we have earlier argued, the $P^\prime$ and $Q^\prime$ fluxes have index structure similar to the geometric flux (namely $\omega_{ij}{}^k$) accompanied by the Levi-Civita tensor, while the $H^\prime$ and $F^\prime$ fluxes have the index structures similar to non-geometric $R^{ijk}$ flux where $\{i, j, k\}$ are the real six-dimensional indices. By this analogy we expect to have the following symplectic components for the S-dual flux pairs $(P^\prime,\, Q^\prime)$ and $(H^\prime,\, F^\prime)$,
\bea
\label{eq:primefluxIIBa}
& & P^\prime_{\alpha\Lambda}, \, {P^\prime}_\alpha{}^\Lambda, \qquad Q^\prime_{\alpha\Lambda}, \, {Q^\prime}_\alpha{}^\Lambda, \qquad H^\prime_{\Lambda}, \, {H^\prime}^\Lambda, \qquad F^\prime_{\Lambda}, \, {F^\prime}^\Lambda\,.
\eea
However, the symplectic pair of flux components which appear in our superpotential have the following respective forms,
\bea
\label{eq:primefluxIIBb}
& & \hskip-1cm P^{\prime\alpha\beta}{}_\Lambda, \, P^{\prime \alpha \beta \Lambda}, \qquad Q^{\prime\alpha\beta}{}_\Lambda, \, Q^{\prime \alpha \beta \Lambda}, \qquad H^{\prime\alpha\beta\gamma}{}_\Lambda, \, H^{\prime \alpha \beta \gamma \Lambda}, \qquad F^{\prime\alpha\beta\gamma}{}_\Lambda, \, F^{\prime \alpha \beta \gamma \Lambda}\,.
\eea
In order to understand the correlation between the two (symplectic/non-symplectic) formulations, now we reconsider the identities given in Eq.~(\ref{eq:Jidentity}), for which we derive the following cohomology formulation:
\bea
\label{eq:Jidentity-cohom}
& & \hskip-1.0cm t^\alpha = \frac{1}{8} \, \ell^{\alpha\beta\gamma}\, \ell_\beta \, \ell_\gamma = \frac{1}{2} \, \ell^{\alpha\beta\gamma}\, \tau_\beta \, \tau_\gamma, \qquad \qquad \\
& & \hskip-1.0cm \, {\cal V} = \frac{1}{8. \, 3!}\, \ell^{\alpha\beta\gamma} \, \ell_\alpha \, \ell_\beta\, \ell_\gamma = \frac{1}{3!}\, \ell^{\alpha\beta\gamma} \, \tau_\alpha \, \tau_\beta\, \tau_\gamma \,, \nonumber
\eea
where $\tau_\alpha$ corresponds to the volume of the 4-cycle and can be written in terms of the 2-cycle volumes as: $\tau_\alpha = \frac{1}{2}\, \ell_{\alpha\beta\gamma}\, t^\beta\, t^\gamma = \frac{1}{2}\, \ell_\alpha$. Here we have used the shorthand notations $\ell_{\alpha} = \ell_{\alpha\beta}\,t^\beta = \ell_{\alpha\beta\gamma}\,t^\beta\, t^\gamma$. Note that, in the absence of odd-axions in our current type IIB construction, $\tau_\alpha = - \, {\rm Im}(T_\alpha)$. In addition, the quantities $\ell^{\alpha\beta\gamma}$ can be defined by using the triple intersections $\ell_{\alpha\beta\gamma}$ as below,
\bea
\label{eq:Inv-lijk}
& & \ell^{\alpha\beta\gamma} = {\cal V}^3\,\ell_{\alpha^\prime\beta^\prime\gamma^\prime}\,{\cal G}^{\alpha\alpha^\prime}\, {\cal G}^{\beta\beta^\prime} \, {\cal G}^{\gamma\gamma^\prime}\,,
\eea
where ${\cal G}^{\alpha\beta}$ denote the inverse moduli space metric defined as under,
\bea
\label{eq:Invmetric-IIB}
& & {\cal G}^{\alpha\beta} = \frac{1}{4\, {\cal V}^2}\, \biggl[ 2\, t^\alpha\, t^\beta - 4\, {\cal V} \, \ell^{\alpha\beta} \biggr]\,.
\eea
This inverse moduli space metric given in Eq.~(\ref{eq:Invmetric-IIB}) leads to an identity ${\cal G}^{\alpha\beta}\, \ell_\beta = 2\, t^\alpha/{\cal V}$ which can be utilized to easily prove our identities given in Eq.~(\ref{eq:Jidentity-cohom}). Using these relations and the inputs from \cite{Aldazabal:2010ef}, we propose that the prime fluxes in Eqs.~(\ref{eq:primefluxIIBa})-(\ref{eq:primefluxIIBb}) are related as:
\bea
\label{eq:primefluxIIBc}
& & {P}^{\prime\beta\gamma{}\Lambda}= {P^\prime}_{\alpha}{}^{\Lambda}\, \ell^{\alpha\beta\gamma}, \qquad \, {P}^{\prime\beta\gamma}{}_{\Lambda} = {P^\prime}_{\alpha \Lambda} \, \ell^{\alpha\beta\gamma}\,,\\
& & {Q}^{\prime\beta\gamma{}\Lambda}= {Q^\prime}_{\alpha}{}^{\Lambda}\, \ell^{\alpha\beta\gamma}, \qquad \, {Q}^{\prime\beta\gamma}{}_{\Lambda} = {Q^\prime}_{\alpha \Lambda} \, \ell^{\alpha\beta\gamma}\,,\nonumber\\
& & {H}^{\prime\alpha\beta\gamma{}\Lambda}= {H^\prime}^{\Lambda}\, \ell^{\alpha\beta\gamma}, \qquad {H}^{\prime\alpha\beta\gamma}{}_{\Lambda} = {H^\prime}_{\Lambda} \, \ell^{\alpha\beta\gamma}\,, \nonumber\\
& & {F}^{\prime\alpha\beta\gamma{}\Lambda}= {F^\prime}^{\Lambda}\, \ell^{\alpha\beta\gamma}, \qquad \, \, {F}^{\prime\alpha\beta\gamma}{}_{\Lambda} = {F^\prime}_{\Lambda} \, \ell^{\alpha\beta\gamma}\,.\nonumber
\eea
Let us illustrate these features by considering an explicit toroidal example, using the orientifold of a ${\mathbb T}^6/({\mathbb Z}_2 \times {\mathbb Z}_2)$ sixfold which has been well studied in the literature. For this setup, we have only one non-zero component of the triple intersection tensor $\ell_{\alpha\beta\gamma}$, namely $\ell_{123} = 1$, and using Eqs.~(\ref{eq:Inv-lijk})-(\ref{eq:Invmetric-IIB}) we find the following simple relations:
\bea
& & \hskip-.5cm \ell_{\alpha\beta\gamma} \, \, \ell^{\alpha\beta\gamma} = \frac{6}{{\cal V}}\, , \qquad \quad \ell_{\alpha\beta\gamma^\prime} \, \, \ell^{\alpha\beta\gamma} = \frac{2\, \delta_{\gamma^\prime}{}^{\gamma}}{{\cal V}}\, , \qquad \quad \ell_{\alpha\beta^\prime\gamma^\prime} \, \, \ell^{\alpha\beta\gamma} = \frac{\delta_{\beta^\prime}{}^\beta\, \delta_{\gamma^\prime}{}^\gamma}{{\cal V}}\,.
\eea
The underlying reason for these relations to hold is the fact that the quantity $\ell^{\alpha\beta\gamma}$ defined in (\ref{eq:Inv-lijk}) takes the following form for this simple toroidal model,
\bea
\label{eq:invkijk-IIB}
& & \ell^{\alpha\beta\gamma} = \frac{\ell_{\alpha\beta\gamma}}{{\cal V}}\,.
\eea
which means that the volume dependence appears only through the overall volume modulus ${\cal V}$, and not in terms of the 4-cycle or 2-cycle volumes. We also note the fact that there is only one non-zero component for the inverse tensor $\ell^{\alpha\beta\gamma}$ which is $\ell^{123} = {\ell_{123}}/{{\cal V}}$. Subsequently, the non-zero components of the various prime fluxes given in Eq.~(\ref{eq:primefluxIIBc}) simplify as,
\bea
\label{eq:primefluxIIBd}
& & {P}^{\prime12{}\Lambda}= {\cal V}^{-1}\, {P^\prime}_{3}{}^{\Lambda}, \quad  {P}^{\prime23{}\Lambda}= {\cal V}^{-1}\, {P^\prime}_{1}{}^{\Lambda}, \quad  {P}^{\prime13{}\Lambda}= {\cal V}^{-1}\, {P^\prime}_{2}{}^{\Lambda}\,,\\
& & {P}^{\prime12}{}_{\Lambda}= {\cal V}^{-1}\, {P^\prime}_{3\, \Lambda}, \quad  {P}^{\prime23}{}_{\Lambda}= {\cal V}^{-1}\, {P^\prime}_{1\, \Lambda}, \quad  {P}^{\prime13}{}_{\Lambda}= {\cal V}^{-1}\, {P^\prime}_{2\, \Lambda}\,,\nonumber\\
& & {Q}^{\prime12{}\Lambda}= {\cal V}^{-1}\, {Q^\prime}_{3}{}^{\Lambda}, \quad  {Q}^{\prime23{}\Lambda}= {\cal V}^{-1}\, {Q^\prime}_{1}{}^{\Lambda}, \quad  {Q}^{\prime13{}\Lambda}= {\cal V}^{-1}\, {Q^\prime}_{2}{}^{\Lambda}\,,\nonumber\\
& & {Q}^{\prime12}{}_{\Lambda}= {\cal V}^{-1}\, {Q^\prime}_{3\, \Lambda}, \quad  {Q}^{\prime23}{}_{\Lambda}= {\cal V}^{-1}\, {Q^\prime}_{1\, \Lambda}, \quad  {Q}^{\prime13}{}_{\Lambda}= {\cal V}^{-1}\, {Q^\prime}_{2\, \Lambda}\,,\nonumber\\
& & {H}^{\prime123{}\Lambda}= {\cal V}^{-1}\, {H^\prime}^{\Lambda}, \quad {H}^{\prime123}{}_{\Lambda} = {\cal V}^{-1}\,{H^\prime}_{\Lambda}, \quad {F}^{\prime123{}\Lambda}= {\cal V}^{-1}\,{F^\prime}^{\Lambda}, \quad {F}^{\prime123}{}_{\Lambda} = {\cal V}^{-1}\, {F^\prime}_{\Lambda}\,,\nonumber
\eea
where $\Lambda \in \{0, 1, 2, 3 \}$. 
This means that we finally have 8 components for each of the fluxes $F, H, H^\prime$ and $F^\prime$ while there are 24 components for each of the $Q, P, P^\prime$ and $Q^\prime$ fluxes. Let us note an important point that the total number of fluxes being 128 corresponds to the $2^{1+h^{1,1}+h^{2,1}}$ which counts the number of generalized flux components of a representation $(2, 2, 2, 2, 2, 2, 2)$ under $SL(2, {\mathbb Z})^7$. Moreover, the correlation between the fluxes as shown in (\ref{eq:primefluxIIBd}) also justifies the earlier appearance of the overall volume ($\cal V$) factor in the toroidal case as have been observed in \cite{Leontaris:2023lfc}, and subsequently Levi-Civita symbol being promoted with the corresponding Levi-Civita tensor in Eq.~(\ref{eq:shortPrimed-flux}). 

Finally, the U-dual completion of the holomorphic flux superpotential having 128 flux components in total along with 7 moduli, namely $\bigl\{S, T_1, T_2, T_3, U^1, U^2, U^3\bigr\}$ for this toroidal model boils down to the following form,
\bea
\label{eq:W-toroidal}
& & \hskip-0.5cm W = \biggl[F_0 + \sum_{i=1}^3 F_i\, U^i + \frac{1}{2} \, \sum_{i\neq j \neq k} \,F^i \,U^j U^k - \, U^1 U^2 U^3 \, F^0 \biggr]\\
& & \hskip0.5cm - \, S \, \biggl[H_0 + \sum_{i=1}^3 H_i\, U^i + \frac{1}{2} \, \sum_{i \neq j \neq k} \, H^i \,U^j U^k - U^1 U^2 U^3 \, H^0 \biggr] \nonumber\\
& & \hskip0.15cm + \sum_{\alpha =1}^3\, T_\alpha\, \biggl[{Q}^\alpha{}_0 + \sum_{i=1}^3 {Q}^\alpha{}_i\, U^i + \frac{1}{2} \, \sum_{i \neq j \neq k} \, {Q}^{\alpha \, i} \,U^j U^k - U^1 U^2 U^3 \, {Q}^{\alpha 0} \biggr]\,\nonumber\\
& & - \, S \sum_{\alpha =1}^3 \, T_\alpha\, \biggl[{P}^\alpha{}_0 + \sum_{i=1}^3 {P}^\alpha{}_i\, U^i + \frac{1}{2} \, \sum_{i \neq j \neq k} \, {P}^{\alpha \, i} \,U^j U^k - U^1 U^2 U^3 \, {P}^{\alpha 0} \biggr]\, \nonumber\\
& & + \frac{1}{2}\, \sum_{\substack{\alpha, \beta =1 \\ \alpha \neq \beta}}^{3} \biggl[T_\alpha T_\beta \biggl\{{P}^{\prime \alpha\beta}{}_{0} + \sum_{i=1}^3 {P}^{\prime \alpha\beta}{}_{i}\, U^i + \frac{1}{2} \, \sum_{i \neq j \neq k} \, {P}^{\prime \alpha\beta i} \, U^j U^k - U^1 U^2 U^3 \, {P}^{\prime \alpha\beta 0} \biggr\}\biggr]\, \nonumber\\
& & - \frac{S}{2}\, \sum_{\substack{\alpha, \beta =1 \\ \alpha \neq \beta}}^{3} \biggl[T_\alpha T_\beta \biggl\{{Q}^{\prime \alpha\beta}{}_{0} + \sum_{i=1}^3 {Q}^{\prime \alpha\beta}{}_{i}\, U^i + \frac{1}{2} \, \sum_{i \neq j \neq k} \, {Q}^{\prime \alpha\beta i}\, U^j U^k - U^1 U^2 U^3\, {Q}^{\prime \alpha\beta 0} \biggr\} \biggr]\, \nonumber\\
& & + \, T_1\, T_2 \, T_3 \biggl[{H}^{\prime123}{}_0 + \sum_{i=1}^3 {H}^{\prime123}{}_i\, U^i + \frac{1}{2} \, \sum_{i \neq j \neq k} \,  {H}^{{\prime123} \, i} \, U^j U^k - U^1 U^2 U^3 \, {H}^{{\prime123} 0} \biggr] \nonumber\\
& & - S\, T_1\, T_2 \, T_3 \biggl[{F}^{\prime123}{}_0 + \sum_{i=1}^3 {F}^{\prime123}{}_i\, U^i + \frac{1}{2} \, \sum_{i \neq j \neq k} \, {F}^{{\prime123} \, i} \,U^j U^k - U^1 U^2 U^3 \,  {F}^{{\prime123} 0}\biggr] \,. \nonumber
\eea

\subsection{Invoking the axionic-flux combinations}
By construction it is clear that after the successive applications of S/T-dualities, the generalized superpotential will have a cubic-form in $T$ and $U$ variables, and a linear form in the axio-dilaton $S$. In what follows, our main goal is to study the insights of the effective four-dimensional scalar potential. 
The generic U-dual completed flux superpotential given in Eq.~(\ref{eq:W-all-2}) can be equivalently written as,
\begin{eqnarray}
\label{eq:W_gen}
& & W = e_\Lambda \, {\cal X}^\Lambda + m^\Lambda \, {\cal F}_\Lambda,
\end{eqnarray} 
where using the flux actions in (\ref{eq:FluxActions-3}) and (\ref{eq:FluxActions-2}), the symplectic vector $(e_\Lambda, m^\Lambda)$ can be given as below,
\begin{eqnarray}
\label{eq:eANDm}
& &  e_\Lambda = \left({F}_\Lambda - S \, {H}_\Lambda\right)  + T_\alpha \left({Q}^{\alpha}{}_\Lambda\,  - S \, {P}^{\alpha}{}_\Lambda \right) + \frac{1}{2} T_\alpha\, T_\beta \left({P}^{\prime\alpha\beta}{}_\Lambda\, - S\, {Q}^{\prime\alpha\beta}{}_\Lambda \right) \, \\
& & \hskip2cm + \, \frac{1}{6} \, T_\alpha \, T_\beta \, T_\gamma \left(H^{\prime\alpha\beta\gamma}{}_\Lambda\, - S \, F^{\prime\alpha\beta\gamma}{}_\Lambda\right), \, \nonumber\\
& &  m^\Lambda = \left({F}^\Lambda - S \, {H}^\Lambda\right)  + T_\alpha \left({Q}^{\alpha}{}^\Lambda\,  - S \, {P}^{\alpha}{}^\Lambda \right) + \frac{1}{2} T_\alpha\, T_\beta \left({P}^{\prime\alpha\beta}{}^\Lambda\, - S\, {Q}^{\prime\alpha\beta}{}^\Lambda \right) \, \nonumber\\
& & \hskip2cm + \, \frac{1}{6} \, T_\alpha \, T_\beta \, T_\gamma \left(H^{\prime\alpha\beta\gamma}{}^\Lambda\, - S \, F^{\prime\alpha\beta\gamma}{}^\Lambda\right).\nonumber
\end{eqnarray} 
Using the superpotential (\ref{eq:W_gen}), one can compute the derivatives with respect to chiral variables, $S$ and $T_\alpha$ which are given as below,
\begin{eqnarray}
\label{eq:W_gen2}
& & W_S = {(e_1)}_\Lambda \, {\cal X}^\Lambda + {(m_1)}^\Lambda \, {\cal F}_\Lambda, \\
& & W_{T_\alpha} = {(e_2)}^\alpha_\Lambda \, {\cal X}^\Lambda + {(m_2)}^\alpha{}^\Lambda \, {\cal F}_\Lambda, \nonumber
\end{eqnarray} 
where the two new pairs of symplectic vectors $(e_1, m_1)$ and $(e_2, m_2)$ are given as: 
\begin{eqnarray}
\label{eq:e1ANDm1}
& &  {(e_1)}_\Lambda = - \biggl[{H}_\Lambda  + T_\alpha \, {P}^{\alpha}{}_\Lambda + \frac{1}{2} T_\alpha\, T_\beta \, {Q}^{\prime\alpha\beta}{}_\Lambda + \, \frac{1}{6} \, T_\alpha \, T_\beta \, T_\gamma \, F^{\prime\alpha\beta\gamma}{}_\Lambda \biggr], \,\\
& &  {(m_1)}^\Lambda = -\biggl[{H}^\Lambda + T_\alpha \, {P}^{\alpha}{}^\Lambda + \frac{1}{2} T_\alpha\, T_\beta \, {Q}^{\prime\alpha\beta}{}^\Lambda  + \, \frac{1}{6} \, T_\alpha \, T_\beta \, T_\gamma  \, F^{\prime\alpha\beta\gamma}{}^\Lambda\, \biggr].\nonumber
\end{eqnarray} 
and
\begin{eqnarray}
\label{eq:e2ANDm2}
& & \hskip-1cm {(e_2)}^\alpha_\Lambda = \left({Q}^{\alpha}{}_\Lambda - S \, {P}^{\alpha}{}_\Lambda \right) + T_\beta \left({P}^{\prime\alpha\beta}{}_\Lambda\, - S\, {Q}^{\prime\alpha\beta}{}_\Lambda \right) + \, \frac{1}{2} T_\beta \, T_\gamma \left(H^{\prime\alpha\beta\gamma}{}_\Lambda\, - S \, F^{\prime\alpha\beta\gamma}{}_\Lambda\right),\\
& & \hskip-1cm {(m_2)}^\alpha{}^\Lambda = \left({Q}^{\alpha}{}^\Lambda\,  - S \, {P}^{\alpha}{}^\Lambda \right) + \, T_\beta \left({P}^{\prime\alpha\beta}{}^\Lambda\, - S\, {Q}^{\prime\alpha\beta}{}^\Lambda \right) + \, \frac{1}{2} \, T_\beta \, T_\gamma \left(H^{\prime\alpha\beta\gamma}{}^\Lambda\, - S \, F^{\prime\alpha\beta\gamma}{}^\Lambda\right).\nonumber
\end{eqnarray} 
Now, we define the following set of the so-called axionic-flux combinations which will turn out to be extremely useful for rearranging the scalar potential pieces into a compact form, 
\bea
\label{eq:AxionicFlux}
& &  {\mathbb H}_\Lambda = {H}_\Lambda  + \rho_\alpha \, {P}^{\alpha}{}_\Lambda + \frac{1}{2} \rho_\alpha\, \rho_\beta \, {Q}^{\prime\alpha\beta}{}_\Lambda + \, \frac{1}{6} \, \rho_\alpha \, \rho_\beta \, \rho_\gamma \, F^{\prime\alpha\beta\gamma}{}_\Lambda\,,\\
& & {\mathbb H}^\Lambda = {H}^\Lambda + \rho_\alpha \, {P}^{\alpha}{}^\Lambda + \frac{1}{2} \rho_\alpha\, \rho_\beta \, {Q}^{\prime\alpha\beta}{}^\Lambda  + \, \frac{1}{6} \, \rho_\alpha \, \rho_\beta \, \rho_\gamma  \, F^{\prime\alpha\beta\gamma}{}^\Lambda\,, \nonumber\\
& &  {\mathbb F}_\Lambda = {F}_\Lambda  + \rho_\alpha \, {Q}^{\alpha}{}_\Lambda + \frac{1}{2} \rho_\alpha\, \rho_\beta \, {P}^{\prime\alpha\beta}{}_\Lambda + \, \frac{1}{6} \, \rho_\alpha \, \rho_\beta \, \rho_\gamma \, H^{\prime\alpha\beta\gamma}{}_\Lambda - \, C_0 \, {\mathbb H}_\Lambda\,, \nonumber\\
& & {\mathbb F}^\Lambda = {F}^\Lambda + \rho_\alpha \, {Q}^{\alpha}{}^\Lambda + \frac{1}{2} \rho_\alpha\, \rho_\beta \, {P}^{\prime\alpha\beta}{}^\Lambda  + \, \frac{1}{6} \, \rho_\alpha \, \rho_\beta \, \rho_\gamma  \, H^{\prime\alpha\beta\gamma}{}^\Lambda\, - \, C_0 \, {\mathbb H}^\Lambda\,, \nonumber
\eea
\bea
& & { {\mathbb P}}^{\alpha}{}_\Lambda = \, {P}^{\alpha}{}_\Lambda + \, \rho_\beta \, {Q}^{\prime\alpha\beta}{}_\Lambda + \, \frac{1}{2} \, \rho_\beta \, \rho_\gamma \, F^{\prime\alpha\beta\gamma}{}_\Lambda\,, \nonumber\\
& & {{\mathbb P}}^{\alpha}{}^\Lambda = {P}^{\alpha}{}^\Lambda +  \rho_\beta \, {Q}^{\prime\alpha\beta}{}^\Lambda  + \, \frac{1}{2} \, \rho_\alpha \, \rho_\beta \, \rho_\gamma  \, F^{\prime\alpha\beta\gamma}{}^\Lambda\,, \nonumber\\
& & { {\mathbb Q}}^{\alpha}{}_\Lambda = \, {Q}^{\alpha}{}_\Lambda + \, \rho_\beta \, {P}^{\prime\alpha\beta}{}_\Lambda + \, \frac{1}{2} \, \rho_\beta \, \rho_\gamma \, H^{\prime\alpha\beta\gamma}{}_\Lambda -\, C_0\, { {\mathbb P}}^{\alpha}{}_\Lambda\,, \nonumber\\
& & {{\mathbb Q}}^{\alpha}{}^\Lambda = {Q}^{\alpha}{}^\Lambda +  \rho_\beta \, {P}^{\prime\alpha\beta}{}^\Lambda  + \, \frac{1}{2} \, \rho_\alpha \, \rho_\beta \, \rho_\gamma  \, H^{\prime\alpha\beta\gamma}{}^\Lambda\, - C_0 \, {{\mathbb P}}^{\alpha}{}^\Lambda\,, \nonumber\\
& & \nonumber\\
& & { {\mathbb Q}}^{\prime\alpha\beta}{}_\Lambda =  {Q}^{\prime\alpha\beta}{}_\Lambda + \, \rho_\gamma \, F^{\prime\alpha\beta\gamma}{}_\Lambda\,, \nonumber\\
& & { {\mathbb Q}}^{\prime\alpha\beta}{}^\Lambda =  {Q}^{\prime\alpha\beta}{}^\Lambda  + \, \rho_\gamma  \, F^{\prime\alpha\beta\gamma}{}^\Lambda\,, \nonumber\\
& & { {\mathbb P}}^{\prime\alpha\beta}{}_\Lambda =  {P}^{\prime\alpha\beta}{}_\Lambda + \, \rho_\gamma \, H^{\prime\alpha\beta\gamma}{}_\Lambda\, - C_0\, { {\mathbb Q}}^{\prime\alpha\beta}{}_\Lambda\,, \nonumber\\
& & { {\mathbb P}}^{\prime\alpha\beta}{}^\Lambda =  {P}^{\prime\alpha\beta}{}^\Lambda  + \, \rho_\gamma  \, H^{\prime\alpha\beta\gamma}{}^\Lambda\, - \, C_0\, { {\mathbb Q}}^{\prime\alpha\beta}{}^\Lambda\,, \nonumber\\
& & \nonumber\\
& & {\mathbb F}^{\prime\alpha\beta\gamma}{}_\Lambda = \, F^{\prime\alpha\beta\gamma}{}_\Lambda\,, \nonumber\\
& & {\mathbb F}^{\prime\alpha\beta\gamma}{}^\Lambda = \, F^{\prime\alpha\beta\gamma}{}^\Lambda\,, \nonumber\\
& & {\mathbb H}^{\prime\alpha\beta\gamma}{}_\Lambda = \, H^{\prime\alpha\beta\gamma}{}_\Lambda\, - \, C_0\,  {\mathbb F}^{\prime\alpha\beta\gamma}{}_\Lambda, \nonumber\\
& & {\mathbb H}^{\prime\alpha\beta\gamma}{}^\Lambda = \, H^{\prime\alpha\beta\gamma}{}^\Lambda\,- C_0\, {\mathbb F}^{\prime\alpha\beta\gamma}{}^\Lambda. \nonumber
\eea
Using these axionic-flux combinations (\ref{eq:AxionicFlux}) along with the definitions of chiral variables in Eq.~(\ref{eq:N=1_coords}), the three pairs of symplectic vectors, namely $(e, m), \, (e_1, m_1)$ and $(e_2, m_2)$ which are respectively given in Eqs.~(\ref{eq:eANDm}), (\ref{eq:e1ANDm1}) and (\ref{eq:e2ANDm2}), can be expressed in the following compact form,
\bea
\label{eq:em}
& & e_\Lambda = \left({\mathbb F}_\Lambda - s \, {\mathbb P}^{}{}_{\Lambda} - {\mathbb P}^{\prime}{}_{\Lambda} + s\, {\mathbb F}^{\prime}{}_{\Lambda} \right)  + i \, \left( - s \, {\mathbb H}_\Lambda - {{\mathbb Q}}^{}{}_{\Lambda} + s \, {{\mathbb Q}}^{\prime}{}_{\Lambda} + {{\mathbb H}}^{\prime}{}_{\Lambda}\right),\\
& & m^\Lambda = \left({\mathbb F}^\Lambda - s \, {\mathbb P}^{}{}^{\Lambda} - {\mathbb P}^{\prime}{}^{\Lambda} + s\, {\mathbb F}^{\prime}{}^{\Lambda} \right)  + i \, \left( -\, s \, {\mathbb H}^\Lambda - {{\mathbb Q}}^{}{}^{\Lambda} + s \, {{\mathbb Q}}^{\prime}{}^{\Lambda} + {{\mathbb H}}^{\prime}{}^{\Lambda}\right), \nonumber
\eea
\bea
\label{eq:e1m1}
& & {(e_1)}_\Lambda = \left(- \, {\mathbb H}_\Lambda + \, {{\mathbb Q}}^{\prime}{}_{\Lambda} \right) + i \left({\mathbb P}^{}{}_{\Lambda} - \, {\mathbb F}^{\prime}{}_{\Lambda} \right)\,,\\
& & {(m_1)}^\Lambda = \left(-\, {\mathbb H}^\Lambda + \, {{\mathbb Q}}^{\prime}{}^{\Lambda}\right) + i \left({\mathbb P}^{}{}^{\Lambda}  - \, {\mathbb F}^{\prime}{}^{\Lambda} \right)\,, \nonumber
\eea
\bea
\label{eq:e2m2}
& & {(e_2)}^\alpha{}_\Lambda = \left({{\mathbb Q}}^{\alpha}{}_{\Lambda} - \, s \, {{\mathbb Q}}^{\prime\alpha}{}_{\Lambda} - \, {{\mathbb H}}^{\prime\alpha}{}_{\Lambda}\right) + i \left(- s \, {\mathbb P}^{\alpha}{}_{\Lambda} - {\mathbb P}^{\prime\alpha}{}_{\Lambda} + s\, {\mathbb F}^{\prime\alpha}{}_{\Lambda} \right)\,,\\
& & {(m_2)}^\alpha{}^\Lambda = \left({{\mathbb Q}}^{\alpha}{}^{\Lambda} - s \, {{\mathbb Q}}^{\prime\alpha}{}^{\Lambda} - {{\mathbb H}}^{\prime\alpha}{}^{\Lambda}\right) + i \left( - \, s \, {\mathbb P}^{\alpha}{}^{\Lambda} - {\mathbb Q}^{\prime\alpha}{}^{\Lambda} + s\, {\mathbb F}^{\prime\alpha}{}^{\Lambda} \right)\,, \nonumber
\eea
where we have used the shorthand notations like ${\mathbb Q}^{}_\Lambda = \tau_\alpha\,{\mathbb Q}^{\alpha}{}_\Lambda$, ${\mathbb Q}^{\prime\alpha}_\Lambda = \,\tau_\beta{\mathbb Q}^{\prime\alpha\beta}{}_\Lambda$, ${\mathbb Q}^{\prime}_\Lambda = \frac12 \tau_\alpha\,\tau_\beta{\mathbb Q}^{\prime\alpha\beta}{}_\Lambda$,  and ${\mathbb H}^{\prime}_\Lambda = \frac16\,\tau_\alpha\tau_\beta \tau_\gamma{\mathbb H}^{\prime\alpha\beta\gamma}{}_\Lambda$ etc. In addition, we mention that such shorthand notations are applicable only with $\tau_\alpha$ contractions, and not to be (conf)used with axionic ($\rho_\alpha$) contractions. This convention will be used wherever the $(Q, P)$, $(P^\prime, Q^\prime)$ and $(H^\prime, F^\prime)$ fluxes are seen with/without a free index $\alpha \in h^{1,1}_+(X)$. Here we recall that $\tau_\alpha = \frac{1}{2}\, {\ell}_{\alpha \beta \gamma} t^\beta t^\gamma$. 


\section{Symplectic formulation of the scalar potential}
\label{sec_potential}
In this section, we will present a compact and concise symplectic formulation for the four-dimensional (effective) scalar potential induced by the generalized fluxes respecting the U-dual completion arguments for the flux superpotential. In our analysis, we start with a superpotential of the form (\ref{eq:W-all-4}) which is more general than the toroidal case. Our approach is to work with the axionic-fluxes  as it helps in simply discarding the explicit presence of the RR ($C_0$ and $C_4$) axions in the game of rewriting the scalar potential in symplectic form, e.g. as seen in \cite{Leontaris:2023lfc}. This, subsequently, also helps us in reducing the number of terms to deal with while working on some explicit construction. We will demonstrate the applicability of our symplectic proposal by considering a simple toroidal model with a flux superpotential resulting in a scalar potential having 76276 pieces while reproducing the same by our master formula.

\subsection{Necessary symplectic identities}
To begin with, let us also recollect some relevant ingredients for rewriting the F-term scalar potential into a symplectic formulation. The strategy we follow is an extension of the previous proposal made in \cite{Shukla:2015hpa}. For the purpose of simplifying the complex structure moduli dependent piece of the scalar potential, we introduce a set of symplectic ingredients. First, we consider the period matrix ${\cal N}$ for the involutively odd (2,1)-cohomology sector which can be expressed using the derivatives of the prepotential as below,
\bea
\label{eq:periodN}
& & {\cal N}_{\Lambda\Delta} = \ov{\cal F}_{\Lambda\Delta} + 2 \, i \, \frac{Im({\cal F}_{\Lambda\Gamma}) \, {\cal X}^\Gamma X^\Sigma \, (Im{\cal F}_{\Sigma \Delta}) }{Im({\cal F}_{\Gamma\Sigma}) {\cal X}^\Gamma X^\Sigma}.
\eea
Subsequently, we define the following hodge star operations acting on the various (odd) three-forms via introducing a set of so-called ${\cal M}$ matrices \cite{Ceresole:1995ca}, 
\begin{eqnarray}
\label{stardef}
&& \star \, {\cal A}_\Lambda =  {\cal M}_{\Lambda}^{\, \, \, \, \, \Sigma} \, \, {\cal A}_\Sigma + {\cal M}_{\Lambda \Sigma}\, \, {\cal B}^\Sigma, \, \, \, {\rm and} \, \, \, \, \star\, {\cal B}^\Lambda = {\cal M}^{\Lambda \Sigma} \, \, {\cal A}_\Sigma + {\cal M}^\Lambda_{\, \, \, \, \Sigma} \, \, {\cal B}^\Sigma\,,
\end{eqnarray}
where
\begin{eqnarray}
\label{coff}
&& {\cal M}^{\Lambda \Delta} = {\rm Im{\cal N}}^{\Lambda \Delta}, \qquad \qquad \qquad {\cal M}_{\Lambda}^{\, \, \, \, \, \Delta}  = {\rm Re{\cal N}}_{\Lambda \Gamma} \, \, {\rm Im{\cal N}}^{\Gamma \Delta}, \\
&& {\cal M}^\Lambda_{\, \, \, \, \Delta} =- \left({\cal M}_{\Lambda}^{\, \, \, \, \, \Delta}\right)^{T} , \qquad \qquad {\cal M}_{\Lambda \Delta}\, =  -{\rm Im{\cal N}}_{\Lambda \Delta} -{\rm Re{\cal N}}_{\Lambda \Sigma} \, \, {\rm Im{\cal N}}^{\Sigma \Gamma}\, \, {\rm Re{\cal N}}_{\Gamma \Delta}. \nonumber
\end{eqnarray}

\subsubsection*{Symplectic identity 1:}
Using the period matrix components, one of the most important identities for simplifying the scalar potential turns out to be the following one \cite{Ceresole:1995ca},
\begin{eqnarray}
\label{eq:Identity1}
&& K^{i \ov j} \, (D_i {\cal X}^{\Lambda}) \, (\ov D_{\ov j} \ov{{\cal X}^{\Delta}}) = - \ov{{\cal X}^{\Lambda}} \,  {\cal X}^{\Delta} - \frac{1}{2} \, e^{-K_{cs}} \, {\rm Im{\cal N}}^{\Lambda \Delta}.
\end{eqnarray}

\subsubsection*{Symplectic identity 2:}
It was observed in \cite{Shukla:2015hpa} that an interesting and very analogous relation as compared to the definition of period matrix (\ref{eq:periodN}) holds which is given as below:
\bea
\label{eq:periodF}
& & {\cal F}_{\Lambda\Delta} = \ov{\cal N}_{\Lambda\Delta} + 2 \, i \, \frac{Im({\cal N}_{\Lambda\Gamma}) \, {\cal X}^\Gamma X^\Sigma \, (Im{\cal N}_{\Sigma \Delta}) }{Im({\cal N}_{\Gamma\Sigma}) {\cal X}^\Gamma X^\Sigma}.
\eea
Moreover, similar to the definition of the period matrices (\ref{coff}), one can also define another set of symplectic quantities given as under,
\begin{eqnarray}
\label{coff2}
&& {\cal L}^{\Lambda \Delta} = {\rm Im{\cal F}}^{\Lambda \Delta},\qquad \qquad \qquad {\cal L}_{\Lambda}^{\, \, \, \, \, \Delta}  = {\rm Re{\cal F}}_{\Lambda \Gamma} \, \, {\rm Im{\cal F}}^{\Gamma \Delta}, \\
&& {\cal L}^\Lambda_{\, \, \, \, \Delta} =- \left({\cal L}_{\Lambda}^{\, \, \, \, \, \Delta}\right)^{T}, \qquad \qquad  {\cal L}_{\Lambda \Delta}\, =  -{\rm Im{\cal F}}_{\Lambda \Delta} -{\rm Re{\cal F}}_{\Lambda \Sigma} \, \, {\rm Im{\cal F}}^{\Sigma \Gamma}\, \, {\rm Re{\cal F}}_{\Gamma \Delta}. \nonumber
\end{eqnarray}

\subsubsection*{Symplectic identity 3:}
The set of ${\cal M}$ and ${\cal L}$ matrices provide another set of very crucial identities given as below,
\bea
\label{eq:symp10}
& & Re({\cal X}^\Lambda \, \ov{\cal X}^\Delta) = -\frac{1}{4} \, e^{-K_{cs}} \, \left( {\cal M}^{\Lambda \Delta} + \, \, {\cal L}^{\Lambda \Delta} \right),
\\
& & Re({\cal X}^\Lambda \, \ov{\cal F}_\Delta) = +\frac{1}{4} \, e^{-K_{cs}} \, \left({\cal M}^\Lambda_{\, \, \, \, \Delta}  +  \, \,  {\cal L}^\Lambda_{\, \, \, \, \Delta} \right), 
\nonumber\\
& & Re({\cal F}_\Lambda \, \ov{\cal X}^\Delta) = - \frac{1}{4} \, e^{-K_{cs}} \, \left(  {\cal M}_{\Lambda}^{\, \, \, \, \, \Delta} +  \, \, {\cal L}_{\Lambda}^{\, \, \, \, \, \Delta}\right), 
\nonumber\\
& & Re({\cal F}_\Lambda \, \ov{\cal F}_\Delta) = + \frac{1}{4} \, e^{-K_{cs}} \, \left({\cal M}_{\Lambda \Delta} + \, \, {\cal L}_{\Lambda \Delta}\right), 
\nonumber
\eea
\bea
\label{eq:symp11}
& & Im({\cal X}^\Lambda \, \ov{\cal X}^\Delta) = + \frac{1}{4} \, e^{-K_{cs}} \, \biggl[\left({\cal M}^{\Lambda}{}_{ \Sigma} \, {\cal L}^{\Sigma \Delta} + {\cal M}^{\Lambda \Sigma} \, {\cal L}_{\Sigma}{}^{ \Delta} \right) \biggr]. 
\\
& & Im({\cal X}^\Lambda \, \ov{\cal F}_\Delta) = - \frac{1}{4} \, e^{-K_{cs}} \, \biggl[\left({\cal M}^{\Lambda}{}_{ \Sigma} \, {\cal L}^{\Sigma}{}_\Delta + {\cal M}^{\Lambda \Sigma} \, {\cal L}_{\Sigma \Delta}\right) - \delta^\Lambda{}_\Delta \biggr]. 
\nonumber\\
& & Im({\cal F}_\Lambda \, \ov{\cal X}^\Delta) = + \frac{1}{4} \, e^{-K_{cs}} \, \biggl[\left({\cal M}_{\Lambda}{}_{ \Sigma} \, {\cal L}^{\Sigma \Delta} + {\cal M}_{\Lambda}{}^{ \Sigma} \, {\cal L}_{\Sigma}{}^{ \Delta} \right) - \delta_\Lambda{}^\Delta\biggr]. 
\nonumber\\
& & Im({\cal F}_\Lambda \, \ov{\cal F}_\Delta) = - \frac{1}{4} \, e^{-K_{cs}} \, \biggl[\left({\cal M}_{\Lambda \Sigma} \, {\cal L}^{\Sigma}{}_{ \Delta} + {\cal M}_{\Lambda}{}^{ \Sigma} \, {\cal L}_{\Sigma \Delta} \right) \biggr].
\nonumber
\eea
Note that the left hand side of these identities are something which explicitly appear in the scalar potential as we will see later on.

\subsubsection*{Symplectic identity 4:}
Apart from the identities given in Eqs.~(\ref{eq:symp10})-(\ref{eq:symp11}), the following non-trivial relations hold which will be more directly useful (as in \cite{Shukla:2015hpa}),
\bea
\label{eq:symp121}
& & \hskip-1.1cm 8 \, e^{K_{cs}} Re({\cal X}^\Gamma \, \ov{\cal X}^\Delta) = {{\cal S}}^{\Gamma}{}_{\Lambda} \left({\cal M}^{\Lambda \Sigma} {{\cal S}}_{\Sigma}^{\, \, \, \, \, \Delta} + {\cal M}^\Lambda_{\, \, \, \, \Sigma} {{\cal S}}^{\Sigma \Delta}\right) - {{\cal S}}^{\Gamma \Lambda} \left({\cal M}_{\Lambda}^{\,\,\, \Sigma} {{\cal S}}_{\Sigma}^{\, \, \, \, \, \Delta} + {\cal M}_{\Lambda \, \Sigma} {{\cal S}}^{\Sigma \Delta}\right) \,, \\
& & \hskip-1.1cm 8 \, e^{K_{cs}} Re({\cal X}^\Gamma \, \ov{\cal F}_\Delta) = {{{\cal S}}}^{\Gamma}{}_{\Lambda} \left({\cal M}^{\Lambda \Sigma} {{\cal S}}_{\Sigma \Delta} + {\cal M}^\Lambda_{\, \, \, \, \Sigma} {{{\cal S}}}^{\Sigma}_{\,\,\,\, \,\,\Delta}\right) - {{{\cal S}}}^{\Gamma \Lambda} \left({\cal M}_{\Lambda}^{\,\,\, \Sigma} {{\cal S}}_{\Sigma \, \Delta} + {\cal M}_{\Lambda \, \Sigma} {{{\cal S}}}^{\Sigma}_{\,\,\,\,\, \Delta}\right) \,, \nonumber\\
& & \hskip-1.1cm 8 \, e^{K_{cs}} Re({\cal F}_\Gamma \, \ov{\cal X}^\Delta) = {{\cal S}}_{\Gamma \Lambda} \left({\cal M}^{\Lambda \Sigma} {{{\cal S}}}_{\Sigma}^{\, \, \, \, \, \Delta} + {\cal M}^\Lambda_{\, \, \, \, \Sigma} {{\cal S}}^{\Sigma \Delta}\right) -{{{\cal S}}}_{\Gamma}{}^{\Lambda} \left({\cal M}_{\Lambda}^{\,\,\, \Sigma} {{{\cal S}}}_{\Sigma}^{\, \, \, \, \, \Delta} + {\cal M}_{\Lambda \, \Sigma} {{\cal S}}^{\Sigma \Delta}\right) \,, \nonumber\\
& & \hskip-1.1cm 8 \, e^{K_{cs}} Re({\cal F}_\Gamma \, \ov{\cal F}_\Delta) = {{\cal S}}_{\Gamma \Lambda} \left({\cal M}^{\Lambda \Sigma} {{\cal S}}_{\Sigma \Delta} + {\cal M}^\Lambda_{\, \, \, \, \Sigma} \, {{{\cal S}}}^{\Sigma}_{\,\,\,\, \,\,\Delta}\right) - {{{\cal S}}}_{\Gamma}{}^{\Lambda} \left({\cal M}_{\Lambda}^{\,\,\, \Sigma} {{\cal S}}_{\Sigma \, \Delta} + {\cal M}_{\Lambda \, \Sigma} \, {{{\cal S}}}^{\Sigma}_{\,\,\,\,\, \Delta}\right) \,,\nonumber
\eea
where
\begin{eqnarray}
\label{coff5}
& & \hskip-1cm  {{\cal S}}^{\Lambda \Delta} = \left({\cal M}^{\Lambda}{}_{ \Sigma} \, {\cal L}^{\Sigma \Delta} + {\cal M}^{\Lambda \Sigma} \, {\cal L}_{\Sigma}{}^{ \Delta} \right), \, \, {{\cal S}}^{\Lambda}{}_{ \Delta}  = - \left({\cal M}^{\Lambda}{}_{ \Sigma} \, {\cal L}^{\Sigma}{}_\Delta + {\cal M}^{\Lambda \Sigma} \, {\cal L}_{\Sigma \Delta}\right) + \delta^\Lambda{}_\Delta, \\
& & \hskip-1cm {{\cal S}}_\Lambda^{\, \, \, \, \Delta} =\left({\cal M}_{\Lambda}{}_{ \Sigma} \, {\cal L}^{\Sigma \Delta} + {\cal M}_{\Lambda}{}^{ \Sigma} \, {\cal L}_{\Sigma}{}^{ \Delta} \right)  - \delta_\Lambda{}^\Delta, \, \, {{\cal S}}_{\Lambda \Delta}\, =  - \left({\cal M}_{\Lambda \Sigma} \, {\cal L}^{\Sigma}{}_{ \Delta} + {\cal M}_{\Lambda}{}^{\Sigma} \, {\cal L}_{\Sigma \Delta} \right). \nonumber
\end{eqnarray}

\subsection{Taxonomy of the scalar potential pieces in three steps}
In the absence of (non-)perturbative corrections, the K\"ahler metric takes a block diagonal form with splitting of pieces coming from generic ${N} = 1$ F-term contribution\footnote{Note that such a splitting of the total scalar potential into two pieces is possible because of the block diagonal nature of the total (inverse) K\"ahler metric. In the absence of odd-moduli $G^a$, the tree level K\"ahler potential is such that there are three blocks corresponding to each of the $S, U^i$ and $T_\alpha$ moduli.},
\begin{eqnarray}
\label{eq:V_gen}
& & \hskip-2cm e^{- K} \, V = K^{{\cal A} \ov {\cal B}} \, (D_{\cal A} W) \, (\ov D_{\ov {\cal B}} \ov{W}) -3 |W|^2 \equiv V_{cs} + V_{k}\, ,
\end{eqnarray}
where
\begin{eqnarray}
\label{eq:VcsVk}
& & \hskip-1.5cm V_{cs} =  K^{{i} \ov {j}} \, (D_{i} W) \, (\ov D_{\ov {j}} \ov{W}), \quad V_{k} =  K^{{A} \ov {B}} \, (D_{A} W) \, (\ov D_{\ov {B}} \ov{W}) -3 |W|^2\,.
\end{eqnarray}
Here, the indices $(i,j)$ correspond to complex structure moduli $U^i$'s while the other indices $(A,B)$ are counted in the remaining chiral variables $\{S, T_\alpha\}$. Using the symplectic identity given in Eq.~(\ref{eq:Identity1}), one can reshuffle the scalar potential pieces $V_{cs}$ and $V_k$ in (\ref{eq:V_gen}) into the following three pieces,
\begin{eqnarray}
& & e^{-K}\, V = V_1 + V_2 + V_3\,,
\end{eqnarray}
where
\begin{eqnarray}
\label{eq:V1V2V3-a}
& & \hskip-1cm V_1 := -\frac{1}{2} \, e^{-K_{cs}} \, \left(e_\Lambda + m^\Sigma \ov {\cal N}_{\Sigma \Lambda} \right) \, {\rm Im{\cal N}}^{\Lambda \Delta} \,   \left(\ov e_\Delta + \ov m^\Gamma {\cal N}_{\Gamma \Delta} \right)\,, \\
& & \hskip-1cm V_2:= - \left(e_\Lambda + m^\Sigma \ov {\cal N}_{\Sigma \Lambda} \right) \, \left(\ov {\cal X}^\Lambda {\cal X}^\Delta \right) \,   \left(\ov e_\Delta + \ov m^\Gamma {\cal N}_{\Gamma \Delta} \right) \nonumber\\
& & + \left(K^{{A} \ov {B}} \, K_A \, K_{\ov {B}} |W|^2 -\,3 |W|^2\right) + K^{{A} \ov {B}} \, \left( (K_A \, W) \, \ov{W}_{\ov B} \, + W_A\, (K_{\ov {B}} \ov W) \right)\,, \nonumber\\
& & \hskip-1cm V_3 := K^{{A} \ov {B}} \, W_A \, \ov{W}_{\ov B} \, .\nonumber
\end{eqnarray}
To appreciate the reason for making such a collection, let us mention that considering the standard GVW superpotential with $H_3/F_3$ fluxes only, one finds that the total scalar potential is entirely contained in the first piece $V_1$ \cite{Taylor:1999ii, Blumenhagen:2003vr}, and $V_2 + V_3$ gets trivial due to the underlying no-scale structure leading to some more internal cancellations. Now, let us recollect some useful relations following from the K\"ahler derivatives and the inverse K\"ahler metric given as below \cite{Grimm:2004uq},
\begin{eqnarray}
\label{eq:derK}
& & K_S = \frac{i}{2 \,s } = - K_{\ov S}, \quad K_{T_\alpha} = -\frac{i \, t^\alpha}{2\cal V} = - K_{\ov{T}_\alpha}\,,
\end{eqnarray}
and
\begin{eqnarray}
\label{eq:InvK}
& & K^{S \ov{S}} =  4 \, s^2, \quad K^{T_\alpha \, \ov{S}} = 0 = K^{S \ov{T}_\alpha}, \quad K^{T_\alpha \, \ov{T}_\beta} = 4\,{\cal G}_{\alpha \beta},
\end{eqnarray}
where we use the following shorthand notations for ${\cal G}$ and ${\cal G}^{-1}$ components,
\begin{eqnarray}
\label{eq:genMetrices}
& & {\cal G}_{\alpha \beta} =  \tau_\alpha \,\tau_\beta - \, {\cal V} \, \ell_{\alpha\beta}, \quad  {\cal G}^{\alpha \beta} = \frac{1}{4{\cal V}^2} \left(2 \, t^\alpha \, t^\beta - \,4 \, {\cal V} \, \ell^{\alpha \beta} \right). 
\end{eqnarray} 
In addition, we have introduced $\ell_0 = 6 {\cal V} = \ell_\alpha\, t^\alpha$, $\ell_\alpha = \ell_{\alpha \beta} \, t^\beta$, and $\ell_{\alpha\beta} = \ell_{\alpha\beta\gamma} \, t^\gamma$. Using the pieces of information in Eqs.~(\ref{eq:derK})-(\ref{eq:InvK}) one gets the following important identities\footnote{These identities hold true for more general cases \cite{AbdusSalam:2020ywo,Cicoli:2017shd}, e.g. in the presence of the perturbative ${\alpha^\prime}^3$-corrections of \cite{Becker:2002nn}, and even when the odd-moduli are included \cite{Cicoli:2021tzt}. However, these relations generically do not hold in the presence of string-loop corrections \cite{Cicoli:2021tzt,Leontaris:2022rzj}.},
\begin{eqnarray}
& & {K}_A\, {K}^{{A} \ov {S}} \,  = (S -\ov S) = - {K}^{{S} \ov {B}} \, {K}_{\ov B}\,, \\
& & {K}_A\, {K}^{{A} \ov {T_\alpha}} \, = (T_\alpha -\ov T_\alpha) = - {K}^{{T_\alpha} \ov {B}} \,  {K}_{\ov B}\,, \nonumber\\
& & K^{{A} \ov {B}} \, K_A \, K_{\ov {B}} = \,4. \nonumber
\end{eqnarray}
Using these identities, the three pieces in Eq.~(\ref{eq:V1V2V3-a}) are further simplified as below,
\begin{eqnarray}
\label{eq:V1V2V3-b}
& & \hskip-1cm V_1 = -\frac{1}{2} \, e^{-K_{cs}} \, \left(e_\Lambda + m^\Sigma \ov {\cal N}_{\Sigma \Lambda} \right) \, {\rm Im{\cal N}}^{\Lambda \Delta} \,   \left(\ov e_\Delta + \ov m^\Gamma {\cal N}_{\Gamma \Delta} \right) \\
& & \hskip-1cm V_2 = - \left(e_\Lambda + m^\Sigma \ov {\cal N}_{\Sigma \Lambda} \right) \, \left(\ov {\cal X}^\Lambda {\cal X}^\Delta \right) \,   \left(\ov e_\Delta + \ov m^\Gamma {\cal N}_{\Gamma \Delta} \right) +|W|^2 \nonumber\\
& &  + (S - \ov{S}) \left(W\, \ov{W}_{\ov S} \, - W_S\, \ov W) \right) + (T_\alpha - \ov{T}_\alpha) \left(W\, \ov{W}_{\ov T_{\alpha}} \, - W_{T_\alpha}\, \ov W) \right) \nonumber\\
& & \hskip-1cm V_3 = 4s^2 \, W_S \, \ov{W}_{\ov S} + 4\, {\cal G}_{\alpha \beta}\, W_{T_\alpha} \, \ov{W}_{\ov T_\beta} \, .\nonumber
\end{eqnarray}
Now, our central goal is to rewrite these three pieces $V_1, V_2$ and $V_3$ in terms of new generalized flux combinations via taking a symplectic approach.

\subsubsection*{Simplifying $V_1$}
Using the S-dual pairs of generalized flux combinations $(e, m)$ as mentioned in Eq.~(\ref{eq:em}), the pieces in $V_1$ can be considered to split into the following two parts,
\bea
\label{eq:Vcs1aAndb}
& & \hskip-0.6cm V_{1} \equiv V_{1}^{(a)} + V_{1}^{(b)} \,,
\eea
where
\bea
\label{eq:Vcs1a}
& & \hskip-0.8cm V_{1}^{(a)} = -\frac{1}{2} \, e^{-K_{cs}} \, \left(e_\Lambda \, {\cal M}^{\Lambda \Delta} \, \ov e_\Delta - e_\Lambda \, {\cal M}^{\Lambda}_{\, \, \, \Delta} \, \ov m^\Delta+ \ov e^\Lambda \, {\cal M}_{\Lambda}^{\, \, \, \Delta} \, m_\Delta- m^\Lambda \, {\cal M}_{\Lambda \Delta} \, \ov m^\Delta\right),
\eea
and
\bea\label{eq:Vcs1b}
& & \hskip-1cm V_{1}^{(b)} = \frac{i}{2} \, e^{-K_{cs}} \,\left(\ov e_\Lambda  m^\Lambda - e_\Lambda  \, \ov m^\Lambda \right).
\eea

\subsubsection*{Simplifying $V_2$}
Similar analysis leads to the following simplifications in the $V_2$ part of the scalar potential,
\bea
\label{eq:Vcs1aAndb}
& & \hskip-0.6cm V_{2} \equiv V_{2}^{(a)} + V_{2}^{(b)} + V_{2}^{(c)} \,,
\eea
where
\bea
& & \hskip-0.3cm V_{2}^{(a)} = |W|^2 - \left(e_\Lambda + m^\Sigma \ov {\cal N}_{\Sigma \Lambda} \right) \, \left(\ov {\cal X}^\Lambda {\cal X}^\Delta \right) \,   \left(\ov e_\Delta + \ov m^\Gamma {\cal N}_{\Gamma \Delta} \right) \\
& & \hskip0.5cm = \left(e_\Lambda \ov e_\Delta - \ov e_\Lambda  e_\Delta \right) \left({\cal X}^\Lambda \,\ov{\cal X}^\Delta \right) + \left(e_\Lambda \ov m^\Delta - \ov e_\Lambda  m^\Delta \right) \left({\cal X}^\Lambda \,\ov{\cal F}_\Delta \right) \nonumber\\
& & \hskip1.5cm + \left(m^\Lambda \ov e_\Delta - \ov m^\Lambda  e_\Delta \right) \left({\cal F}_\Lambda \,\ov{\cal X}^\Delta \right) + \left(m^\Lambda \ov m^\Delta - \ov m^\Lambda  m^\Delta \right) \left({\cal F}_\Lambda \,\ov{\cal F}_\Delta \right) \nonumber\\
& & \hskip-0.3cm V_{2}^{(b)} = (S - \ov{S}) \left(W\, \ov{W}_{\ov S} \, - W_S\, \ov W) \right) \nonumber\\
& & \hskip0.5cm = (2\, i\, s) \biggl[\left(e_\Lambda \ov {(e_1)}_\Delta -  {(e_1)}_\Lambda  \ov e_\Delta \right) \left({\cal X}^\Lambda \,\ov{\cal X}^\Delta \right) + \quad \dots \quad + \quad \dots \quad + \quad \dots \biggr]\,, \nonumber\\
& & \hskip-0.3cm V_{2}^{(c)} = (T_\alpha - \ov{T}_\alpha) \left(W\, \ov{W}_{\ov T_{\alpha}} \, - W_{T_\alpha}\, \ov W) \right) \nonumber\\
& & \hskip0.5cm = (-2\, i\, \tau_\alpha) \biggl[\left(e_\Lambda \ov {(e_2)^\alpha}_\Delta -  {(e_2)^\alpha}_\Lambda  \ov e_\Delta \right) \left({\cal X}^\Lambda \,\ov{\cal X}^\Delta \right) + \quad \dots \quad + \quad \dots \quad + \quad \dots \biggr]\,. \nonumber
\eea
As explicitly mentioned in the second line of the piece $V_{2}^{(a)}$, here $\dots$ denotes the analogous pieces involving $\left({\cal X}^\Lambda \,\ov{\cal F}_\Delta \right)$, $\left({\cal F}_\Lambda \,\ov{\cal X}^\Delta \right)$ and $\left({\cal F}_\Lambda \,\ov{\cal F}_\Delta \right)$ and having the flux indices being appropriately contracted.

\subsubsection*{Simplifying $V_3$}
Now considering the inverse K\"ahler metric in Eq.~(\ref{eq:InvK}) along with derivatives of the superpotential using the new generalized flux orbits in Eq.~(\ref{eq:AxionicFlux}), one gets the following rearrangement of $V_3$ after a very painstaking reshuffling of the various pieces,
\begin{eqnarray}
\label{eq:V3final}
& & \hskip-0.6cm V_3 = 4s^2 \, W_S \, \ov{W}_{\ov S} + 4\, {\cal G}_{\alpha \beta}\, W_{T_\alpha} \, \ov{W}_{\ov T_\beta} \\
& & \hskip0cm = 4\, s^2 \,  \biggl[{(e_1)}_\Lambda \ov {(e_1)}_\Delta \left({\cal X}^\Lambda \,\ov{\cal X}^\Delta \right) + \quad \dots \quad + \quad \dots \quad + \quad \dots \biggr]\,, \nonumber\\
& & \hskip0cm  + \, 4\, {\cal G}_{\alpha \beta}\, \biggl[{(e_2)^\alpha}_\Lambda \ov {(e_2)^\beta}_\Delta \left({\cal X}^\Lambda \,\ov{\cal X}^\Delta \right) + \quad \dots \quad + \quad \dots \quad + \quad \dots \biggr]\,, \nonumber
\end{eqnarray}
It is worth mentioning again at this point that the rearrangement of terms using new versions of the ``generalized flux orbits" have been performed in an iterative manner, in a series of papers \cite{Blumenhagen:2013hva, Gao:2015nra, Shukla:2015rua, Shukla:2015bca, Shukla:2015hpa}, which have set some guiding rules for next step intuitive generalization, otherwise the rearrangement even at the intermediate steps is very peculiar and it could be much harder to directly arrive at a final form without earlier motivations. 

The full scalar potential can be expressed in 36 types of terms such that there are 20 of those which are of (${\cal O}_1 \wedge \ast {\cal O}_2$) type, while the remaining 16 terms are of (${\cal O}_1 \wedge {\cal O}_2$) type where ${\cal O}_1$ and ${\cal O}_2$ can denote some real function of fluxes and axions. As a particular toroidal case, these can simply be the standard 128 fluxes or their respective axionic-flux combinations as we will discuss in a moment. The generic the scalar potential arising from the U-dual completed flux superpotential can be expressed as below,
\bea
\label{eq:symplectic-1a}
& & V = V_{{\cal O}_1 \wedge \ast {\cal O}_2} + V_{{\cal O}_1 \wedge {\cal O}_2}, 
\eea
where
\bea
\label{eq:symplectic-1b}
& & \hskip-0.5cm V_{{\cal O}_1 \wedge \ast {\cal O}_2} = V_{\mathbb F \mathbb F} + V_{\mathbb H \mathbb H}+ V_{\mathbb Q \mathbb Q}+ V_{\mathbb P \mathbb P}+ V_{{\mathbb P}^\prime {\mathbb P}^\prime} + V_{{\mathbb Q}^\prime {\mathbb Q}^\prime} + V_{{\mathbb H}^\prime {\mathbb H}^\prime}+ V_{{\mathbb F}^\prime {\mathbb F}^\prime} \\
& & \hskip1.5cm + V_{\mathbb F \mathbb P}+ V_{\mathbb F {\mathbb P}^\prime}+ V_{\mathbb F {\mathbb F}^\prime}+ V_{\mathbb H \mathbb Q}+ V_{\mathbb H {\mathbb Q}^\prime}+ V_{\mathbb H {\mathbb H}^\prime}+ V_{\mathbb Q {\mathbb Q}^\prime}+ V_{\mathbb Q{\mathbb H}^\prime} \nonumber\\
& & \hskip1.5cm + V_{\mathbb P {\mathbb P}^\prime}+ V_{\mathbb P {\mathbb F}^\prime}+ V_{{\mathbb P}^\prime {\mathbb F}^\prime}+ V_{{\mathbb Q}^\prime {\mathbb H}^\prime},  \nonumber\\
& & \hskip-0.5cm V_{{\cal O}_1 \wedge {\cal O}_2} = V_{\mathbb F \mathbb H}+ V_{\mathbb F \mathbb Q}+ V_{\mathbb F {\mathbb Q}^\prime}+ V_{\mathbb F {\mathbb H}^\prime} + V_{\mathbb H \mathbb P}+ V_{\mathbb H {\mathbb P}^\prime}+ V_{\mathbb H {\mathbb F}^\prime} \nonumber\\
& & \hskip1.5cm + V_{\mathbb Q \mathbb P}+ V_{\mathbb Q {\mathbb P}^\prime}+ V_{\mathbb Q {\mathbb F}^\prime} + V_{\mathbb P {\mathbb Q}^\prime} + V_{\mathbb P {\mathbb H}^\prime} \nonumber\\
& & \hskip1.5cm  + V_{{\mathbb P}^\prime {\mathbb Q}^\prime}+ V_{{\mathbb P}^\prime {\mathbb H}^\prime} + V_{{\mathbb Q}^\prime {\mathbb F}^\prime} + V_{{\mathbb H}^\prime {\mathbb F}^\prime}\,. \nonumber
\eea
While we give the full details about each of these 36 terms in the Appendix \ref{sec_Appendix1}, let us mention a couple insights about our master formula (\ref{eq:symplectic-1a})-(\ref{eq:symplectic-1b}):
\begin{itemize}

\item
In the absence of prime fluxes 26 pieces of the scalar potential are projected out and there remain only 10 pieces, namely $\left\{V_{\mathbb F \mathbb F}, V_{\mathbb H \mathbb H}, V_{\mathbb Q \mathbb Q}, V_{\mathbb P \mathbb P}, V_{\mathbb H \mathbb Q}, V_{\mathbb F \mathbb P} \right\}$ which are of  (${\cal O}_1 \wedge \ast {\cal O}_2$) type, and $\left\{V_{\mathbb F \mathbb H}, V_{\mathbb F \mathbb Q}, V_{\mathbb P \mathbb Q}\right\}$ which are of (${\cal O}_1 \wedge {\cal O}_2$) type\footnote{Recall that the axionic-flux combinations involved in these 10 terms generically depend on prime indexed fluxes as well and therefore explicit expressions of these axionic-flux combinations will simplify in their absence. Therefore, it should not be naively assumed that the internal structure of these 10 pieces remain the same in the absence of prime fluxes.}. This is what has been presented in \cite{Gao:2015nra, Shukla:2016hyy}. Let us note that the analysis of the scalar potential in \cite{Gao:2015nra} was performed by using the internal background metric of the toroidal model, and a generalisation to symplectic formulation was proposed in \cite{Shukla:2016hyy} which bypasses the need of knowing the metric for the internal manifold via using symplectic ingredients along with moduli space metrics on the K\"ahler- and complex structure-moduli dependent sectors.

\item
The S-duality among the various pieces of the scalar potential has been also manifested from our collection. For example, we have the following S-dual invariant pieces among the overall 36 pieces of the scalar potential:
\bea
\label{eq:potential-S-dual-pairs}
& & \hskip-0.5cm \left(V_{\mathbb F \mathbb F} + V_{\mathbb H \mathbb H}\right), \quad \left(V_{\mathbb Q \mathbb Q} + V_{\mathbb P \mathbb P}\right), \quad \left(V_{{\mathbb P}^\prime {\mathbb P}^\prime} + V_{{\mathbb Q}^\prime {\mathbb Q}^\prime}\right), \quad \left(V_{{\mathbb H}^\prime {\mathbb H}^\prime} + V_{{\mathbb F}^\prime {\mathbb F}^\prime}\right), \quad \left(V_{\mathbb F \mathbb P} + V_{\mathbb H \mathbb Q}\right),\\
& & \hskip-0.5cm \left(V_{\mathbb F {\mathbb P}^\prime} + V_{\mathbb H {\mathbb Q}^\prime}\right), \quad \left(V_{\mathbb F {\mathbb F}^\prime} + V_{\mathbb H {\mathbb H}^\prime}\right), \quad \left(V_{\mathbb Q {\mathbb Q}^\prime} + V_{\mathbb P {\mathbb P}^\prime}\right), \quad \left(V_{\mathbb Q {\mathbb H}^\prime} + V_{\mathbb P {\mathbb F}^\prime}\right), \quad \left(V_{{\mathbb P}^\prime {\mathbb F}^\prime} + V_{{\mathbb Q}^\prime {\mathbb H}^\prime}\right), \nonumber\\
& & \hskip-0.5cm \left(V_{\mathbb F \mathbb Q} + V_{\mathbb H \mathbb P}\right), \quad \left(V_{\mathbb F {\mathbb Q}^\prime} + V_{\mathbb H {\mathbb P}^\prime}\right), \quad \left(V_{\mathbb F {\mathbb H}^\prime} + V_{\mathbb H {\mathbb F}^\prime}\right), \quad \left(V_{\mathbb Q {\mathbb P}^\prime} + V_{\mathbb P {\mathbb Q}^\prime}\right), \quad \left(V_{\mathbb Q {\mathbb F}^\prime} + V_{\mathbb P {\mathbb H}^\prime}\right), \nonumber\\
& & \hskip-0.5cm \left(V_{{\mathbb P}^\prime {\mathbb H}^\prime} + V_{{\mathbb Q}^\prime {\mathbb F}^\prime}\right), \quad \left(V_{\mathbb F \mathbb H} \right),\quad \left(V_{\mathbb Q \mathbb P} \right), \quad \left(V_{{\mathbb P}^\prime {\mathbb Q}^\prime}\right), \quad \left(V_{{\mathbb H}^\prime {\mathbb F}^\prime}\right). \nonumber
\eea
Assuming that the complex structure moduli as well as the Einstein-frame volume moduli do not transform under the S-duality operations, one can easily verify the above-mentioned claims by using the following transformations,
\bea
& & s \to \frac{s}{s^2+ C_0^2}, \qquad C_0 \to -\frac{C_0}{s^2+ C_0^2}, \qquad \frac{C_0}{s} \to -\frac{C_0}{s}, \qquad \rho_\alpha \to \rho_\alpha,  
\eea
As a quick-check, one can consider the case of standard GVW superpotential with $(F, H)$ fluxes only, then we have
\bea
& & V_{\mathbb F \mathbb F} + V_{\mathbb H \mathbb H} \simeq \frac{F \wedge \ast F}{s} + \frac{s^2 + C_0^2}{s} \, H \wedge \ast H - \frac{C_0}{s} (F \wedge \ast H + H \wedge \ast F)
\eea
Given that $\{F \to H, H \to -F\}$ under S-duality, the first two pieces are S-dual to each other while the last piece being a product of two anti S-dual pieces is self S-dual. In this way, our symplectic formulation can be considered to be in a manifestly S-duality invariant form as one can see it explicitly with some little efforts.

\item
It is well understood that all the pieces of ${\cal O}_1 \wedge \ast {\cal O}_2$ type involve the information about the internal metric while working in the so-called standard formulation based on the real six-dimensional indices (e.g. see \cite{Gao:2015nra,Leontaris:2023lfc}), and therefore cannot appear as a topological term. On the other hand, pieces of ${\cal O}_1 \wedge  {\cal O}_2$ type can usually appear as tadpole contributions. However, let us note that in the presence of non-geometric fluxes, especially the non-geometric S-dual pair of fluxes, $ {\cal O}_1 \wedge {\cal O}_2$ may not be entirely a tadpole piece though it may have a tadpole like term within it \cite{Leontaris:2023lfc}. For example, even in the absence of prime fluxes, the $V_{{\mathbb Q} {\mathbb P}}$ piece has some information about the internal background via period/metric inputs which can also be observed from the non-symplectic formulation of the scalar potential \cite{Gao:2015nra} in which such a piece explicitly involves the internal metric implying that the $V_{{\mathbb P} {\mathbb Q}}$ piece is not topological. 

\end{itemize}


\subsection{Master formula}

Although we have presented all the 36 types of flux-bilinear pieces possible in the scalar potential, the attempts so far have just been to elaborate on the insights of various terms and how they could appear from the flux superpotential in connection with the standard U-dual flux parameters, and it is desirable that we club these 36 terms in a more concise symplectic formulation. Aiming at this goal we investigated the 36 pieces in some more detail and managed to rewrite the full scalar potential in just a few terms of $({\cal O}_1 \wedge \ast \ov {\cal O}_2)$ and $({\cal O}_1 \wedge \ov {\cal O}_2)$ types as we express below,
\bea
\label{eq:symplectic-2}
& & V =  V_{({\cal O}_1 \wedge \ast \ov {\cal O}_2)} + V_{({\cal O}_1 \wedge \ov {\cal O}_2)},
\eea
where
\bea
\label{eq:symplectic-3}
& &  \hskip-1.5cm V_{({\cal O}_1 \wedge \ast \ov {\cal O}_2)} = -\frac{1}{4\,s\, {\cal V}^2}\,  \int_{X_6} \biggl[\chi \wedge \ast \ov\chi + {\widetilde\psi} \wedge \ast {\ov{\widetilde\psi}} +  {\cal G}_{\alpha\beta} \, {\widetilde\Psi^\alpha} \wedge \ast \ov{\widetilde\Psi^\beta} \\
& & \hskip0.75cm + \frac{i}{2}\left(\widetilde{\chi} \wedge \ast  {\ov{\widetilde\psi}} - \ov{\widetilde{\chi}} \wedge \ast  {\widetilde\psi} \right) + \frac{i}{2}\left(\widetilde\Psi \wedge \ast \ov{\widetilde{\chi}} - \ov{\widetilde\Psi} \wedge \ast \widetilde{\chi} \right)  \biggr], \nonumber\\
& & \hskip-1.5cm V_{({\cal O}_1 \wedge \ov {\cal O}_2)} = -\frac{1}{4\,s\, {\cal V}^2}\,  \int_{X_6} \biggl[ (-i) \left(\chi \wedge \ov\chi + \chi \wedge \ov{\widetilde\chi} + 2 \, {\widetilde\psi} \wedge {\ov{\widetilde\psi}} + 2\,{\cal G}_{\alpha\beta} \,{\Psi^\alpha} \wedge \ov{\widetilde\Psi^\beta} \right)\nonumber\\
& & \hskip0.75cm + \left(\widetilde{\chi} \wedge  {\ov{\widetilde\psi}} + \ov{\widetilde{\chi}} \wedge {\widetilde\psi} \right) + \left(\widetilde\Psi \wedge \ov{\widetilde{\chi}} + \ov{\widetilde\Psi} \wedge \widetilde{\chi} \right) \biggr]. \nonumber
\eea
The compact formulation given in Eqs.~(\ref{eq:symplectic-2})-(\ref{eq:symplectic-3}) involves only three types of complex axionic-flux combinations which are generically defined as,
\bea
\label{eq:chi-psi-Psi}
& & {\rm Flux} = \, {\rm Flux}^\Lambda \, {\cal A}_\Lambda + {\rm Flux}_\Lambda \, {\cal B}^\Lambda,
\eea
where the symbol ``Flux" in the above denotes ${\rm Flux} = \{\chi, \psi, \Psi\}$ and electric/magnetic components of these fluxes are given in terms of the axionic-flux combinations as below,
\bea
\label{eq:psi}
& & \psi_\Lambda = s\, \left(- \, {\mathbb H}_\Lambda + \, {{\mathbb Q}}^{\prime}{}_{\Lambda} \right) + i \,  s \, \left({\mathbb P}^{}{}_{\Lambda} - \, {\mathbb F}^{\prime}{}_{\Lambda} \right)\,,\\
& & \psi^\Lambda = s\, \left(-\, {\mathbb H}^\Lambda + \, {{\mathbb Q}}^{\prime}{}^{\Lambda}\right) + i \, s \, \left({\mathbb P}^{}{}^{\Lambda}  - \, {\mathbb F}^{\prime}{}^{\Lambda} \right)\,, \nonumber
\eea
\bea
\label{eq:chi}
& & \chi_\Lambda = \left({\mathbb F}_\Lambda - {\mathbb P}^{\prime}{}_{\Lambda} \right)  + i \, \left(- {{\mathbb Q}}^{}{}_{\Lambda} + {{\mathbb H}}^{\prime}{}_{\Lambda}\right) + i\, \psi_\Lambda,\\
& & \chi^\Lambda = \left({\mathbb F}^\Lambda - {\mathbb P}^{\prime}{}^{\Lambda} \right)  + i \, \left(- {{\mathbb Q}}^{}{}^{\Lambda} + {{\mathbb H}}^{\prime}{}^{\Lambda}\right) + i\,  \psi^\Lambda, \nonumber
\eea
\bea
\label{eq:Psi}
& & \Psi^\alpha{}_\Lambda = \left({{\mathbb Q}}^{\alpha}{}_{\Lambda} - \, s \, {{\mathbb Q}}^{\prime\alpha}{}_{\Lambda} - \, {{\mathbb H}}^{\prime\alpha}{}_{\Lambda}\right) + i \left(- s \, {\mathbb P}^{\alpha}{}_{\Lambda} - {\mathbb P}^{\prime\alpha}{}_{\Lambda} + s\, {\mathbb F}^{\prime\alpha}{}_{\Lambda} \right)\,,\\
& & \Psi^\alpha{}^\Lambda = \left({{\mathbb Q}}^{\alpha}{}^{\Lambda} - s \, {{\mathbb Q}}^{\prime\alpha}{}^{\Lambda} - {{\mathbb H}}^{\prime\alpha}{}^{\Lambda}\right) + i \left( - \, s \, {\mathbb P}^{\alpha}{}^{\Lambda} - {\mathbb Q}^{\prime\alpha}{}^{\Lambda} + s\, {\mathbb F}^{\prime\alpha}{}^{\Lambda} \right)\,. \nonumber
\eea
As we have argued earlier, here use the shorthand notations like ${\mathbb Q}^{}_\Lambda = \tau_\alpha\,{\mathbb Q}^{\alpha}{}_\Lambda$, ${\mathbb Q}^{\prime\alpha}_\Lambda = \,\tau_\beta{\mathbb Q}^{\prime\alpha\beta}{}_\Lambda$, ${\mathbb Q}^{\prime}_\Lambda = \frac12 \tau_\alpha\,\tau_\beta{\mathbb Q}^{\prime\alpha\beta}{}_\Lambda$,  and ${\mathbb H}^{\prime}_\Lambda = \frac16\,\tau_\alpha\tau_\beta \tau_\gamma{\mathbb H}^{\prime\alpha\beta\gamma}{}_\Lambda$ etc. In the similar way we write $\Psi_\Lambda = \tau_\alpha \, \Psi^\alpha{}_\Lambda$ and $\Psi^\Lambda = \tau_\alpha\, \Psi^\alpha{}^\Lambda$ wherever $\Psi$ appears without an $h^{1,1}_+$ index $\alpha$. Subsequently, we will have the following relations consistent with out shorthand notations,
\bea
\label{eq:Psi-2}
& & \Psi{}_\Lambda = \left({{\mathbb Q}}^{}{}_{\Lambda} - 2\, s \, {{\mathbb Q}}^{\prime}{}_{\Lambda} - 3\, {{\mathbb H}}^{\prime}{}_{\Lambda}\right) + i \left(- s \, {\mathbb P}^{}{}_{\Lambda} - 2\, {\mathbb P}^{\prime}{}_{\Lambda} + 3\,s\, {\mathbb F}^{\prime}{}_{\Lambda} \right)\,,\\
& & \Psi{}^\Lambda = \left({{\mathbb Q}}^{}{}^{\Lambda} - 2\,s \, {{\mathbb Q}}^{\prime}{}^{\Lambda} - 3\, {{\mathbb H}}^{\prime}{}^{\Lambda}\right) + i \left( - \, s \, {\mathbb P}^{}{}^{\Lambda} - 2\, {\mathbb Q}^{\prime}{}^{\Lambda} + 3\, s\, {\mathbb F}^{\prime}{}^{\Lambda} \right)\,. \nonumber
\eea
In addition, the so-called tilde fluxes for $\chi, \psi$ and $\Psi^\alpha$ are defined as below,
\bea
\label{eq:tilde-chi-psi-Psi}
& & \widetilde{\chi} = -\left({{\cal S}}^{\Sigma\Delta} \chi{}_\Delta + {{\cal S}}^\Sigma{}_\Delta \chi^{\Delta}\right) {\cal A}_\Sigma + \left({{\cal S}}_{\Sigma}{}^{\Delta} \chi{}_\Delta + {{\cal S}}_{\Sigma\Delta} \chi^{\Delta}\right) {\cal B}^\Sigma, \\
& & \widetilde{\psi} = -\left({{\cal S}}^{\Sigma\Delta} \psi{}_\Delta + {{\cal S}}^\Sigma{}_\Delta \psi^{\Delta}\right) {\cal A}_\Sigma + \left({{\cal S}}_{\Sigma}{}^{\Delta} \psi{}_\Delta + {{\cal S}}_{\Sigma\Delta} \psi^{\Delta}\right) {\cal B}^\Sigma, \nonumber\\
& & \widetilde{\Psi^\alpha} = -\left({{\cal S}}^{\Sigma\Delta} \Psi^\alpha{}_\Delta + {{\cal S}}^\Sigma{}_\Delta \Psi^\alpha{}^{\Delta}\right) {\cal A}_\Sigma + \left({{\cal S}}_{\Sigma}{}^{\Delta} \Psi^\alpha{}_\Delta + {{\cal S}}_{\Sigma\Delta} \Psi^\alpha{}^{\Delta}\right) {\cal B}^\Sigma. \nonumber
\eea 
Now we demonstrate the use of our symplectic formulation, in particular the master formula (\ref{eq:symplectic-2})-(\ref{eq:symplectic-3}) by considering an explicit (toroidal) example.

\subsection{Demonstrating the formulation for an explicit example}
In order to demonstrate our symplectic proposal for an explicit example, we again get back to our friend, the toroidal type IIB model based on $\mathbb T^6/(\mathbb Z_2 \times \mathbb Z_2)$ orientifold. As a particular case, several scenarios can be considered by switching-off certain fluxes at a time. We have performed a detailed analysis of all the 36 terms of the scalar potential in full generality, as collected in the Appendix \ref{sec_Appendix1}. This leads to a total of 76276 terms while being expressed in terms of the usual fluxes, however this number reduces to 10888 terms when the total scalar potential is expressed in terms of the axionic-fluxes (\ref{eq:AxionicFlux}), subsequently leading to the numerics about the number of terms in each of the 36 pieces as presented in Table \ref{tab_term-counting}. Moreover, in light of recovering the results from our master formula (\ref{eq:symplectic-2})-(\ref{eq:symplectic-3}), we have a clear splitting of 10888 terms in the following manner,
\bea
& & \#\left(V_{({\cal O}_1 \wedge \ast \ov {\cal O}_2)}\right) = 5576, \qquad  \#\left(V_{({\cal O}_1 \wedge \ov {\cal O}_2)}\right) = 5312.
\eea
To appreciate the importance of the axionic-flux polynomials we present Table \ref{tab_term-counting}.
\begin{table}[h!]
\begin{center}
\begin{tabular}{|c||c|c||c|c|} 
\hline
& &&&\\
& Fluxes & $\#$(terms) in $V$ using & Axionic-fluxes & $\#$(terms) in $V$ using \\
& & standard fluxes & & axionic-fluxes in (\ref{eq:AxionicFlux}) \\
\hline
& &&&\\
1 & $F$  & 76 & ${\mathbb F}$ & 76  \\
& & & & \\
& & & & \\
2 & $F, H$ & 361 & ${\mathbb F}, {\mathbb H}$ & 160  \\
& & & & \\
& & & & \\
3 & $F, H, Q$ & 2422 & ${\mathbb F}, {\mathbb H}, {\mathbb Q}$ & 772  \\
& &&&\\
& &&&\\
4 & $F, H, Q, P$ & 9661 & ${\mathbb F}, {\mathbb H}, {\mathbb Q}, {\mathbb P}$ & 2356  \\
& &&&\\
& &&&\\
5 & $F, H, Q, P, P'$  & 23314 & ${\mathbb F}, {\mathbb H}, {\mathbb Q}, {\mathbb P}, {\mathbb P}'$ & 4855  \\
& &&&\\
& &&&\\
6 & $F, H, Q, P$  & 50185 & ${\mathbb F}, {\mathbb H}, {\mathbb Q}, {\mathbb P},$ & 8326  \\
& $P', Q'$ & &  ${\mathbb P}', {\mathbb Q}'$ & \\
& &&&\\
7 & $F, H, Q, P,$  & 60750 & ${\mathbb F}, {\mathbb H}, {\mathbb Q}, {\mathbb P},$ & 9603  \\
& $P', Q', H'$ & &  ${\mathbb P}', {\mathbb Q}', {\mathbb H}'$ &\\
& & & & \\
8 & $F, H, Q, P, $  & 76276 & ${\mathbb F}, {\mathbb H}, {\mathbb Q}, {\mathbb P},$ & 10888  \\
& $P', Q', H', F'$ & & ${\mathbb P}', {\mathbb Q}', {\mathbb H}', {\mathbb F}'$ &\\
\hline
\end{tabular}
\end{center}
\caption{Counting of scalar potential terms for a set of fluxes being turned-on at a time.}
\label{tab_term-counting}
\end{table}


\section{Summary and conclusions}
\label{sec_conclusions}

The U-dual completion of the flux superpotential in the type IIB supergravity theory leads to the inclusion of four pairs of S-dual fluxes which has attracted some significant amount of interest in the recent past \cite{Aldazabal:2006up,Aldazabal:2008zza,Aldazabal:2010ef,Lombardo:2016swq,Lombardo:2017yme}. This idea of the U-dual completion of the flux superpotential has been mostly studied in the context of toroidal setting using an orientifold of a $\mathbb T^6/(\mathbb Z_2 \times \mathbb Z_2)$ orbifold. In this regard, some interesting insights of this flux superpotential have been recently explored from the point of view of the four-dimensional scalar potential in \cite{Leontaris:2023lfc} where the full scalar potential has been reformulated in terms of the metric of the internal toroidal sixfold. 

Given that the analytic expression for the metric of a generic CY threefold is not known, in order to promote this U-dual completion arguments beyond the toroidal cases one needs to rewrite the scalar potential in a symplectic formulation. In this article, we have filled this gap by presenting a symplectic master-formula for the four-dimensional $N =1$ scalar potential induced by a generalized superpotential with U-dual fluxes. For this purpose, first we invoked the symplectic version of the prime fluxes introduced in (\ref{eq:All-flux}) by taking lessons from the toroidal constructions in \cite{Aldazabal:2006up,Aldazabal:2008zza,Aldazabal:2010ef,Lombardo:2016swq,Lombardo:2017yme,Leontaris:2023lfc}. In this process we derived the cohomology formulation of two important identities (\ref{eq:Jidentity}) which were useful in establishing the connection between the Heterotic compactification models and the type IIB setup having U-dual fluxes in \cite{Aldazabal:2010ef}, and we present this identity in Eq.~(\ref{eq:Jidentity-cohom}). In the second step, we have invoked the so-called axionic-fluxes, collected in Eq.~(\ref{eq:AxionicFlux}), which are some specific combinations of RR axions $(C_2/C_4)$  and the fluxes to be directly used in rewriting the scalar potential pieces summarized in the appendix \ref{sec_Appendix1}. 

Finally, using the 36 pieces as presented in the appendix \ref{sec_Appendix1} we construct a compact and concise version of the generic scalar potential in the form of following master-formula which is written in terms of three axionic-flux combinations, namely $\chi, \psi$ and $\Psi$ being defined in Eqs.~(\ref{eq:psi})-(\ref{eq:Psi}),
\bea
\label{eq:symplectic-4}
& & \hskip-1.5cm V =  -\frac{1}{4\,s\, {\cal V}^2}\, \int_{X_6} \biggl[\chi \wedge \ast \ov\chi + {\widetilde\psi} \wedge \ast {\ov{\widetilde\psi}} +  {\cal G}_{\alpha\beta} \, {\widetilde\Psi^\alpha} \wedge \ast \ov{\widetilde\Psi^\beta} \\
& & \hskip0.5cm + \frac{i}{2}\left(\widetilde{\chi} \wedge \ast  {\ov{\widetilde\psi}} - \ov{\widetilde{\chi}} \wedge \ast  {\widetilde\psi} \right) + \frac{i}{2}\left(\widetilde\Psi \wedge \ast \ov{\widetilde{\chi}} - \ov{\widetilde\Psi} \wedge \ast \widetilde{\chi} \right), \nonumber\\
& & \hskip0.5cm + (-i) \left(\chi \wedge \ov\chi + \chi \wedge \ov{\widetilde\chi} + 2 \, {\widetilde\psi} \wedge {\ov{\widetilde\psi}} + 2\,{\cal G}_{\alpha\beta} \,{\Psi^\alpha} \wedge \ov{\widetilde\Psi^\beta} \right)\nonumber\\
& & \hskip0.5cm + \,\left(\widetilde{\chi} \wedge  {\ov{\widetilde\psi}} + \ov{\widetilde{\chi}} \wedge {\widetilde\psi} \right) + \left(\widetilde\Psi \wedge \ov{\widetilde{\chi}} + \ov{\widetilde\Psi} \wedge \widetilde{\chi} \right) \biggr]. \nonumber
\eea
This master-formula is generically valid for models beyond the toroidal constructions, and can be considered as a generalization of a series of works presented in \cite{Blumenhagen:2013hva, Gao:2015nra, Shukla:2015rua, Shukla:2015bca, Shukla:2015hpa,Blumenhagen:2015lta,Shukla:2016hyy,Shukla:2019wfo,Shukla:2019dqd,Shukla:2019akv,Shukla:2019wfo}. Finally, in order to demonstrate the utility of the master-formula we have re-derived the results of \cite{Leontaris:2023lfc} by recovering all the 76276 terms of the scalar potential induced via a generalized flux superpotential. It would be interesting to understand if this scalar potential can arise from a more fundamental framework such as some S-dual completion of the Double Field Theory on the lines of \cite{Blumenhagen:2013hva,Shukla:2015hpa,Blumenhagen:2015lta}. It will also be interesting to perform a detailed study of the Bianchi identities and the tadpole cancellation conditions in this symplectic formulation. We hope to get back to addressing some of these issues in a future work.


\section*{Acknowledgments}
We would like to thank Gerardo Aldazabal, Sayan Biswas, Ralph Blumenhagen, Xin Gao, Mariana Gra$\tilde{n}$a, Fernando Marchesano, Fabio Riccioni, Wieland Staessens and Rui Sun for useful discussions (on related topics) on various different occasions in the past. PS would like to thank the {\it Department of Science and Technology (DST), India} for the kind support.


\appendix
\setcounter{equation}{0}


\section{Collection of various scalar potential pieces}
\label{sec_Appendix1}
The first collection of the scalar potential pieces as mentioned in Eq.~(\ref{eq:symplectic-1b}) has all the 20 terms of the type (${\cal O}_1 \wedge \ast {\cal O}_2$), while the second collection has 16 terms of (${\cal O}_1 \wedge {\cal O}_2$) type. Now we present the explicit and detailed forms of the 36 scalar potential pieces. 

\subsection{Pieces of (${\cal O}_1 \wedge \ast {\cal O}_2$) type}
These (${\cal O}_1 \wedge \ast {\cal O}_2$) type of pieces can be further classified into what we call as the ``diagonal-pieces" and the ``cross-pieces". Using $e^{\cal Y} =  - \frac{1}{4\, s\, {\cal V}^2}$, we will express such terms as below:

\subsubsection*{Diagonal-pieces :}
\bea
\label{eq:diag-1a}
& (1): & \quad V_{\mathbb F \mathbb F} = e^{\cal Y} \, \int_{X_6} \,{\mathbb F} \wedge \ast {\mathbb F},\\
& (2): & \quad V_{\mathbb H \mathbb H} = e^{\cal Y} \, \int_{X_6} \, s^2 \, \, {\mathbb H} \wedge \ast {\mathbb H},  \nonumber\\
& (3): & \quad V_{{\mathbb Q} {\mathbb Q}} = e^{\cal Y} \, \int_{X_6} \left( {\mathbb Q} \wedge \ast {\mathbb Q} - \widetilde{\cal Q} \wedge \ast \widetilde{\cal Q} + {\cal G}_{\alpha\beta}\, \, \widetilde{\cal Q}^\alpha \wedge \ast \widetilde{\cal Q}^\beta\right) ,  \nonumber\\
& (4): & \quad V_{{\mathbb P} {\mathbb P}} = e^{\cal Y} \, \int_{X_6}  s^2 \,\left( {\mathbb P} \wedge \ast {\mathbb P} - \widetilde{\cal P} \wedge \ast \widetilde{\cal P} + {\cal G}_{\alpha\beta}\, \, \widetilde{\cal P}^\alpha \wedge \ast \widetilde{\cal P}^\beta\right), \nonumber\\
& (5): & \quad V_{{\mathbb P}^\prime {\mathbb P}^\prime} = e^{\cal Y} \, \int_{X_6} \left( {\mathbb P}^\prime \wedge \ast {\mathbb P}^\prime - 2\, \widetilde{\cal P}^\prime \wedge \ast \widetilde{\cal P}^\prime + {\cal G}_{\alpha\beta}\, \, \widetilde{\cal P}^{\prime\alpha} \wedge \ast \widetilde{\cal P}^{\prime\beta}\right), \nonumber\\
& (6): & \quad V_{{\mathbb Q}^\prime {\mathbb Q}^\prime} = e^{\cal Y} \, \int_{X_6} s^2 \,\left( {\mathbb Q}^\prime \wedge \ast {\mathbb Q}^\prime - 2\, \widetilde{\cal Q}^\prime \wedge \ast \widetilde{\cal Q}^\prime + {\cal G}_{\alpha\beta}\, \, \widetilde{\cal Q}^{\prime\alpha} \wedge \ast \widetilde{\cal Q}^{\prime\beta}\right) ,  \nonumber\\
& (7): & \quad V_{{\mathbb H}^\prime {\mathbb H}^\prime} = e^{\cal Y} \, \int_{X_6} \left( {\mathbb H}^\prime \wedge \ast {\mathbb H}^\prime - 3\, \widetilde{\cal H}^\prime \wedge \ast \widetilde{\cal H}^\prime + {\cal G}_{\alpha\beta}\, \, \widetilde{\cal H}^{\prime\alpha} \wedge \ast \widetilde{\cal H}^{\prime\beta}\right), \nonumber\\
& (8): & \quad V_{{\mathbb F}^\prime {\mathbb F}^\prime} = e^{\cal Y} \, \int_{X_6} s^2 \,\left( {\mathbb F}^\prime \wedge \ast {\mathbb F}^\prime - 3\, \widetilde{\cal F}^\prime \wedge \ast \widetilde{\cal F}^\prime + {\cal G}_{\alpha\beta}\, \, \widetilde{\cal F}^{\prime\alpha} \wedge \ast \widetilde{\cal F}^{\prime\beta}\right).  \nonumber
\eea
Here we have used the following definitions for the so-called ``tilde" fluxes corresponding to a given (axionic)flux, as proposed/used in Eq.~(\ref{eq:tilde-chi-psi-Psi}),
\bea
& & \widetilde{\cal Q}^\alpha = -\left({{\cal S}}^{\Sigma\Delta} \hat{\mathbb Q}^\alpha{}_\Delta + {{\cal S}}^\Sigma{}_\Delta \hat{\mathbb Q}^{\alpha\Delta}\right) {\cal A}_\Sigma + \left({{\cal S}}_{\Sigma}{}^{\Delta} \hat{\mathbb Q}^\alpha{}_\Delta + {{\cal S}}_{\Sigma\Delta} \hat{\mathbb Q}^{\alpha\Delta}\right) {\cal B}^\Sigma, \\
& & \widetilde{\cal P}^\alpha = -\left({{\cal S}}^{\Sigma\Delta} \hat{\mathbb P}^\alpha{}_\Delta + {{\cal S}}^\Sigma{}_\Delta \hat{\mathbb P}^{\alpha\Delta}\right) {\cal A}_\Sigma + \left({{\cal S}}_{\Sigma}{}^{\Delta} \hat{\mathbb P}^\alpha{}_\Delta + {{\cal S}}_{\Sigma\Delta} \hat{\mathbb P}^{\alpha\Delta}\right) {\cal B}^\Sigma,\nonumber
\eea 
and $\widetilde{\cal Q} = \widetilde{\cal Q}^\alpha \tau_\alpha$, $\widetilde{\cal P} = \widetilde{\cal P}^\alpha \tau_\alpha$, $\widetilde{\cal P}^\prime = \frac{1}{2}\widetilde{\cal P}^{\prime\alpha\beta} \tau_\alpha\tau_\beta$, $\widetilde{\cal F}^\prime = \frac{1}{6}\widetilde{\cal F}^{\prime\alpha\beta\gamma} \tau_\alpha\tau_\beta\tau_\gamma$ etc. 

\noindent
It maybe worth to mention that for the toroidal case we have the following relations,
\bea
& & \int_{X_6} \left({\cal G}_{\alpha\beta}\, \, \widetilde{\cal H}^{\prime\alpha} \wedge \ast \widetilde{\cal H}^{\prime\beta} - 3\, \widetilde{\cal H}^\prime \wedge \ast \widetilde{\cal H}^\prime \right) = 0, \\
& & \int_{X_6} \left( {\cal G}_{\alpha\beta}\, \, \widetilde{\cal F}^{\prime\alpha} \wedge \ast \widetilde{\cal F}^{\prime\beta} - 3\, \widetilde{\cal F}^\prime \wedge \ast \widetilde{\cal F}^\prime \right) = 0.  \nonumber
\eea

\subsubsection*{Cross-pieces :}
\bea
\label{eq:cross1a}
& (9): & \quad V_{\mathbb F {\mathbb P}} = e^{\cal Y} \, \int_{X_6} (-2 \,s) \, \left({\mathbb F} \wedge \ast {\mathbb P}  - \widetilde{\cal F} \wedge \ast \widetilde{\cal P}\right), \\
& (10): & \quad V_{\mathbb F {\mathbb P}^\prime} = e^{\cal Y} \, \int_{X_6} (-2) \, \left({\mathbb F} \wedge \ast {\mathbb P}^\prime  - \widetilde{\cal F} \wedge \ast \widetilde{\cal P}^\prime\right), \nonumber\\
& (11): & \quad V_{\mathbb F {\mathbb F}^\prime} = e^{\cal Y} \, \int_{X_6} (2\,s) \, \left({\mathbb F} \wedge \ast {\mathbb F}^\prime  - 2\, \widetilde{\cal F} \wedge \ast \widetilde{\cal F}^\prime\right), \nonumber\\
& (12): & \quad V_{\mathbb H {\mathbb Q}} = e^{\cal Y} \, \int_{X_6} (2 \,s) \, \left({\mathbb H} \wedge \ast {\mathbb Q}  - \widetilde{\cal H} \wedge \ast \widetilde{\cal Q}\right),\nonumber\\
& (13): & \quad V_{\mathbb H {\mathbb Q}^\prime} = e^{\cal Y} \, \int_{X_6} (-2\,s^2) \, \left({\mathbb H} \wedge \ast {\mathbb Q}^\prime  - \widetilde{\cal H} \wedge \ast \widetilde{\cal Q}^\prime\right), \nonumber\\
& (14): & \quad V_{\mathbb H {\mathbb H}^\prime} = e^{\cal Y} \, \int_{X_6} (-2\,s) \, \left({\mathbb H} \wedge \ast {\mathbb H}^\prime  - 2\, \widetilde{\cal H} \wedge \ast \widetilde{\cal H}^\prime\right), \nonumber\\
& (15): & \quad V_{\mathbb Q {\mathbb Q}^\prime} = e^{\cal Y} \, \int_{X_6} (-2\,s) \,\left( {\mathbb Q} \wedge \ast {\mathbb Q}^\prime - 2\, \widetilde{\cal Q} \wedge \ast \widetilde{\cal Q}^\prime + {\cal G}_{\alpha\beta}\, \, \widetilde{\cal Q}^{\alpha} \wedge \ast \widetilde{\cal Q}^{\prime\beta}\right) ,  \nonumber\\
& (16): & \quad V_{\mathbb Q {\mathbb H}^\prime} = e^{\cal Y} \, \int_{X_6} (-2) \,\left( {\mathbb Q} \wedge \ast {\mathbb H}^\prime - 2\, \widetilde{\cal Q} \wedge \ast \widetilde{\cal H}^\prime + {\cal G}_{\alpha\beta}\, \, \widetilde{\cal Q}^{\alpha} \wedge \ast \widetilde{\cal H}^{\prime\beta}\right) ,  \nonumber\\
& (17): & \quad V_{\mathbb P {\mathbb P}^\prime} = e^{\cal Y} \, \int_{X_6} (2\,s) \,\left( {\mathbb P} \wedge \ast {\mathbb P}^\prime - 2\, \widetilde{\cal P} \wedge \ast \widetilde{\cal P}^\prime + {\cal G}_{\alpha\beta}\, \, \widetilde{\cal P}^{\alpha} \wedge \ast \widetilde{\cal P}^{\prime\beta}\right) ,  \nonumber\\
& (18): & \quad V_{\mathbb P {\mathbb F}^\prime} = e^{\cal Y} \, \int_{X_6} (-2\,s^2) \,\left( {\mathbb P} \wedge \ast {\mathbb F}^\prime - 2\, \widetilde{\cal P} \wedge \ast \widetilde{\cal F}^\prime + {\cal G}_{\alpha\beta}\, \, \widetilde{\cal P}^{\alpha} \wedge \ast \widetilde{\cal F}^{\prime\beta}\right) ,  \nonumber\\
& (19): & \quad V_{{\mathbb P}^\prime {\mathbb F}^\prime} = e^{\cal Y} \, \int_{X_6} (-2\,s) \,\left({\mathbb P}^\prime \wedge \ast {\mathbb F}^\prime - 3\, \widetilde{\cal P}^\prime \wedge \ast \widetilde{\cal F}^\prime + {\cal G}_{\alpha\beta}\, \, \widetilde{\cal P}^{\prime\alpha} \wedge \ast \widetilde{\cal F}^{\prime\beta}\right),  \nonumber\\
& (20): & \quad V_{{\mathbb Q}^\prime {\mathbb H}^\prime} = e^{\cal Y} \, \int_{X_6} (2\,s) \,\left({\mathbb Q}^\prime \wedge \ast {\mathbb H}^\prime - 3\, \widetilde{\cal Q}^\prime \wedge \ast \widetilde{\cal H}^\prime + {\cal G}_{\alpha\beta}\, \, \widetilde{\cal Q}^{\prime\alpha} \wedge \ast \widetilde{\cal H}^{\prime\beta}\right)\, , \nonumber
\eea

\subsection{Pieces of (${\cal O}_1 \wedge {\cal O}_2$) type}

\bea
\label{eq:cross1b}
& (21): & \quad V_{\mathbb F \mathbb H} = e^{\cal Y} \, \int_{X_6} (2\,s) \, \, \mathbb F \wedge \mathbb H, \\
& (22): & \quad V_{\mathbb F {\mathbb Q}} = e^{\cal Y} \, \int_{X_6} (2) \, \, \mathbb F \wedge {\mathbb Q}, \nonumber\\
& (23): & \quad V_{\mathbb F {\mathbb Q}^\prime} = e^{\cal Y} \, \int_{X_6} (-2\,s) \, \left(\mathbb F \wedge {\mathbb Q}^\prime - \mathbb F \wedge \widetilde{\cal Q}^\prime - \widetilde{\cal F} \wedge {\mathbb Q}^\prime \right), \nonumber\\
& (24): & \quad V_{\mathbb F {\mathbb H}^\prime} = e^{\cal Y} \, \int_{X_6} (-2) \, \left(\mathbb F \wedge {\mathbb H}^\prime - \mathbb F \wedge \widetilde{\cal H}^\prime - \widetilde{\cal F} \wedge {\mathbb H}^\prime \right), \nonumber
\eea
\bea
& (25): & \quad V_{\mathbb H {\mathbb P}} = e^{\cal Y} \, \int_{X_6} (2\, s^2) \, \, {\mathbb H} \wedge {\mathbb P}, \nonumber\\
& (26): & \quad V_{\mathbb H {\mathbb P}^\prime} = e^{\cal Y} \, \int_{X_6} (2\,s) \, \left(\mathbb H \wedge {\mathbb P}^\prime - \mathbb H \wedge \widetilde{\cal P}^\prime - \widetilde{\cal H} \wedge {\mathbb P}^\prime \right), \nonumber\\
& (27): & \quad V_{\mathbb H {\mathbb F}^\prime} = e^{\cal Y} \, \int_{X_6} (-2\, s^2) \, \left(\mathbb H \wedge {\mathbb F}^\prime - \mathbb H \wedge \widetilde{\cal F}^\prime - \widetilde{\cal H} \wedge {\mathbb F}^\prime \right), \nonumber\\
& (28): & \quad V_{{\mathbb Q} {\mathbb P}} = e^{\cal Y} \, \int_{X_6} (2\,s) \biggl[{\mathbb Q} \wedge {\mathbb P} - \left({\mathbb Q} \wedge \widetilde{\cal P} + \widetilde{\cal Q} \wedge {\mathbb P}\right)  +{\cal G}_{\alpha\beta} \left({\mathbb Q}^\alpha \wedge \widetilde{\cal P}^\beta + \widetilde{\cal Q}^\alpha \wedge {\mathbb P}^\beta \right) \biggr], \nonumber\\
& (29): & \quad V_{\mathbb Q {\mathbb P}^\prime} = e^{\cal Y} \, \int_{X_6} (2) \biggl[{\mathbb Q} \wedge {\mathbb P}^\prime - \left({\mathbb Q} \wedge \widetilde{\cal P}^\prime + \widetilde{\cal Q} \wedge {\mathbb P}^\prime\right)  +{\cal G}_{\alpha\beta} \left({\mathbb Q}^\alpha \wedge \widetilde{\cal P}^{\prime\beta} + \widetilde{\cal Q}^\alpha \wedge {\mathbb P}^{\prime\beta} \right) \biggr], \nonumber\\
& (30): & \quad V_{\mathbb Q {\mathbb F}^\prime} = e^{\cal Y} \, \int_{X_6} (-2\,s) \biggl[{\mathbb Q} \wedge {\mathbb F}^\prime - 2\left({\mathbb Q} \wedge \widetilde{\cal F}^\prime + \widetilde{\cal Q} \wedge {\mathbb F}^\prime\right)  +{\cal G}_{\alpha\beta} \left({\mathbb Q}^\alpha \wedge \widetilde{\cal F}^{\prime\beta} + \widetilde{\cal Q}^\alpha \wedge {\mathbb F}^{\prime\beta} \right) \biggr], \nonumber\\
& (31): & \quad V_{\mathbb P {\mathbb Q}^\prime} = e^{\cal Y} \, \int_{X_6} (2\,s^2) \biggl[{\mathbb P} \wedge {\mathbb Q}^\prime - \left({\mathbb P} \wedge \widetilde{\cal Q}^\prime + \widetilde{\cal P} \wedge {\mathbb Q}^\prime\right)  +{\cal G}_{\alpha\beta} \left({\mathbb P}^\alpha \wedge \widetilde{\cal Q}^{\prime\beta} + \widetilde{\cal P}^\alpha \wedge {\mathbb Q}^{\prime\beta} \right) \biggr], \nonumber\\
& (32): & \quad V_{\mathbb P {\mathbb H}^\prime} = e^{\cal Y} \, \int_{X_6} (2\,s) \biggl[{\mathbb P} \wedge {\mathbb H}^\prime - 2\left({\mathbb P} \wedge \widetilde{\cal H}^\prime + \widetilde{\cal P} \wedge {\mathbb H}^\prime\right)  +{\cal G}_{\alpha\beta} \left({\mathbb P}^\alpha \wedge \widetilde{\cal H}^{\prime\beta} + \widetilde{\cal P}^\alpha \wedge {\mathbb H}^{\prime\beta} \right) \biggr], \nonumber\\
& (33): & \quad V_{{\mathbb P}^\prime {\mathbb Q}^\prime} = e^{\cal Y} \, \int_{X_6} (2\,s) \biggl[{\mathbb P}^\prime \wedge {\mathbb Q}^\prime - 2\left({\mathbb P}^\prime \wedge \widetilde{\cal Q}^\prime + \widetilde{\cal P}^\prime \wedge {\mathbb Q}^\prime\right)  +{\cal G}_{\alpha\beta} \left({\mathbb P}^{\prime\alpha} \wedge \widetilde{\cal Q}^{\prime\beta} + \widetilde{\cal P}^{\prime\alpha} \wedge {\mathbb Q}^{\prime\beta} \right) \biggr], \nonumber\\
& (34): & \quad V_{{\mathbb P}^\prime {\mathbb H}^\prime} = e^{\cal Y} \, \int_{X_6} (2) \biggl[{\mathbb P}^\prime \wedge {\mathbb H}^\prime - 2\left({\mathbb P}^\prime \wedge \widetilde{\cal H}^\prime + \widetilde{\cal P}^\prime \wedge {\mathbb H}^\prime\right)  +{\cal G}_{\alpha\beta} \left({\mathbb P}^{\prime\alpha} \wedge \widetilde{\cal H}^{\prime\beta} + \widetilde{\cal P}^{\prime\alpha} \wedge {\mathbb H}^{\prime\beta} \right) \biggr], \nonumber\\
& (35): & \quad V_{{\mathbb Q}^\prime {\mathbb F}^\prime} = e^{\cal Y} \, \int_{X_6} (2\,s^2) \biggl[{\mathbb Q}^\prime \wedge {\mathbb F}^\prime - 2\left({\mathbb Q}^\prime \wedge \widetilde{\cal F}^\prime + \widetilde{\cal Q}^\prime \wedge {\mathbb F}^\prime\right)  +{\cal G}_{\alpha\beta} \left({\mathbb Q}^{\prime\alpha} \wedge \widetilde{\cal F}^{\prime\beta} + \widetilde{\cal Q}^{\prime\alpha} \wedge {\mathbb F}^{\prime\beta} \right) \biggr], \nonumber\\
& (36): & \quad V_{{\mathbb H}^\prime {\mathbb F}^\prime} = e^{\cal Y} \, \int_{X_6} (2\,s) \biggl[{\mathbb H}^\prime \wedge {\mathbb F}^\prime - 3\left({\mathbb H}^\prime \wedge \widetilde{\cal F}^\prime + \widetilde{\cal H}^\prime \wedge {\mathbb F}^\prime\right)  +{\cal G}_{\alpha\beta} \left({\mathbb H}^{\prime\alpha} \wedge \widetilde{\cal F}^{\prime\beta} + \widetilde{\cal H}^{\prime\alpha} \wedge {\mathbb F}^{\prime\beta} \right) \biggr]. \nonumber
\eea


\bibliographystyle{utphys}
\bibliography{reference}

\end{document}